\documentclass{aa}
\usepackage[varg]{txfonts}
\usepackage{graphicx}
\usepackage{t1enc}
\usepackage{color}
\usepackage{amsmath,esint}

\newcommand{\kms}   {km~s$^{-1}$}
\newcommand{\kmspc} {km~s$^{-1}$ pc$^{-1}$}
\newcommand{\Msol}  {$M_{\sun}$}
\newcommand{\Msolpc}{$M_{\sun}$ pc$^{-1}$}
\newcommand{\Lsol}  {$L_{\sun}$}
\newcommand{\nh}    {NH$_3$}
\newcommand{\cmd}   {cm$^{-2}$}
\newcommand{\cmt}   {cm$^{-3}$}
\newcommand{\hho}   {H$_2$O}
\newcommand{\hh}    {H$_2$}
\newcommand{\mjyb}  {mJy~beam$^{-1}$}
\newcommand{\hco}   {HCO$^+$}   
\newcommand{\halpha}{H$\alpha$}  
\newcommand{\Herschel}{\textit{Herschel}}
\newcommand{\sii}   {[\ion{S}{ii}]}
\newcommand{\T}[1]  {\ensuremath{T_\mathrm{#1}}}  
\newcommand{\Tbg}   {\T{bg}}
\newcommand{\Tex}   {\T{ex}}  
  
\newcommand{\Trot}  {\T{rot}}  
\newcommand{\Ml}    {\ensuremath{M_\mathrm{lin}}}
\newcommand{\Mlc}   {\ensuremath{M_\mathrm{lin,crit}}}
\newcommand{\RA}[4] {$\alpha(J2000)=#1^\mathrm{h}#2^\mathrm{m}#3\fs#4$}
\newcommand{\DECf}[4]{$\delta(J2000)=#1\degr#2'#3\farcs#4$}
\newcommand{\DEC}[3]{$\delta(J2000)=#1\degr#2'#3''$}
\newcommand{\equal}{\ensuremath{\!\!\!&=&\!\!\!}}

\begin{document}

\title{ 
VLA ammonia observations of L1287: analysis of\\  
the ``Guitar'' core and two filaments}
\titlerunning{Ammonia observations of L1287}

\author{
Inma~Sep\'ulveda\inst{1} \and 
Robert~Estalella\inst{1} \and 
Guillem~Anglada\inst{2}  \and 
Rosario~L\'opez\inst{1}  \and 
Angels~Riera\inst{3}\thanks{Deceased on 2017 September 27}     \and \\
Gemma~Busquet\inst{4,5} \and
Aina~Palau\inst{6}      \and
Jos\'e M. Torrelles\inst{4,5} \and
Luis F. Rodr\'{\i}guez\inst{6}
}
\institute{
Departament de F\'{\i}sica Qu\`antica i Astrof\'{\i}sica,
Institut de Ci\`encies del Cosmos, Universitat de Barcelona, (IEEC-UB)
Mart\'{i} i Franqu\`es, 1, E-08028 Barcelona, Spain. 
\and 
Instituto de Astrof\'{\i}sica de Andaluc\'{\i}a, CSIC, 
Glorieta de la Astronom\'{\i}a, s/n, E-18008 Granada, Spain.
\and 
Departament de F\'{\i}sica i Enginyeria Nuclear, Universitat Polit\`ecnica de
Catalunya, Av.\ Eduard Maristany, 16, E-08019 Barcelona, Catalunya, Spain.
\and 
Institut de Ci\`encies de l'Espai (ICE, CSIC), Campus UAB,  
Carrer de Can Magrans, s/n, E-08193, Cerdanyola del Vall\`es, Catalunya, Spain.
\and
Institut d'Estudis Espacials de Catalunya (IEEC), E-08034, Barcelona, Catalunya, Spain.
\and 
Instituto de Radioastronom\'{\i}a y Astrof\'{\i}sica, Universidad Nacional 
Aut\'onoma de M\'exico, P.O. Box 3-72, 58090, Morelia, Michoac\'an, Mexico
}
\authorrunning{Sep\'ulveda et al.}

\date{
Received /
Accepted 
}

\abstract{}{
The present work aims at studying the dense gas of the molecular cloud LDN 1287 (L1287), which harbors a double FU Ori system, an energetic molecular outflow and a still-forming cluster of deeply embedded low-mass, young stellar objects, showing a high level of fragmentation.
}{
We present optical H$\alpha$ and \sii{}, and VLA \nh{} $(1,1)$ and $(2,2)$ observations with an angular resolution of $\sim3\farcs5$. The observed \nh{} spectra have been analyzed with the HfS tool, fitting simultaneously three different velocity components.
}{
The \nh{} emission from L1287 comes from four different structures: a core associated with RNO 1, a guitar-shaped core (the ``Guitar'') and two interlaced filaments (the Blue and Red Filaments) roughly centered towards the binary FU Ori system RNO 1B/C and its associated cluster. 
Regarding the Guitar Core, there are clear signatures of gas infall onto a central mass that has been estimated to be $\sim2.1$ \Msol. 
Regarding the two filaments, they have radii $\sim0.03$ pc,  masses per unit length $\sim50$ \Msolpc, and are near isothermal equilibrium. 
A central cavity, probably related with the outflow, and also traced by the H$\alpha$ and \sii{} emission,  is identified, with several young stellar objects near its inner walls. 
Both filaments show clear signs of perturbation by the high-velocity gas of the outflows driven by one or several young stellar objects of the cluster. 
The Blue and Red filaments are coherent in velocity and have nearly subsonic gas motions, except at the position of the embedded sources. 
Velocity gradients across the Blue Filament can be interpreted either as infalling material onto the filament or rotation. 
Velocity gradients along the filaments are interpreted as infall motions towards a gravitational well at the intersection of both filaments
}{}


\maketitle  

\section{Introduction}

\begin{figure*}[htb]
\resizebox{\hsize}{!}{\includegraphics{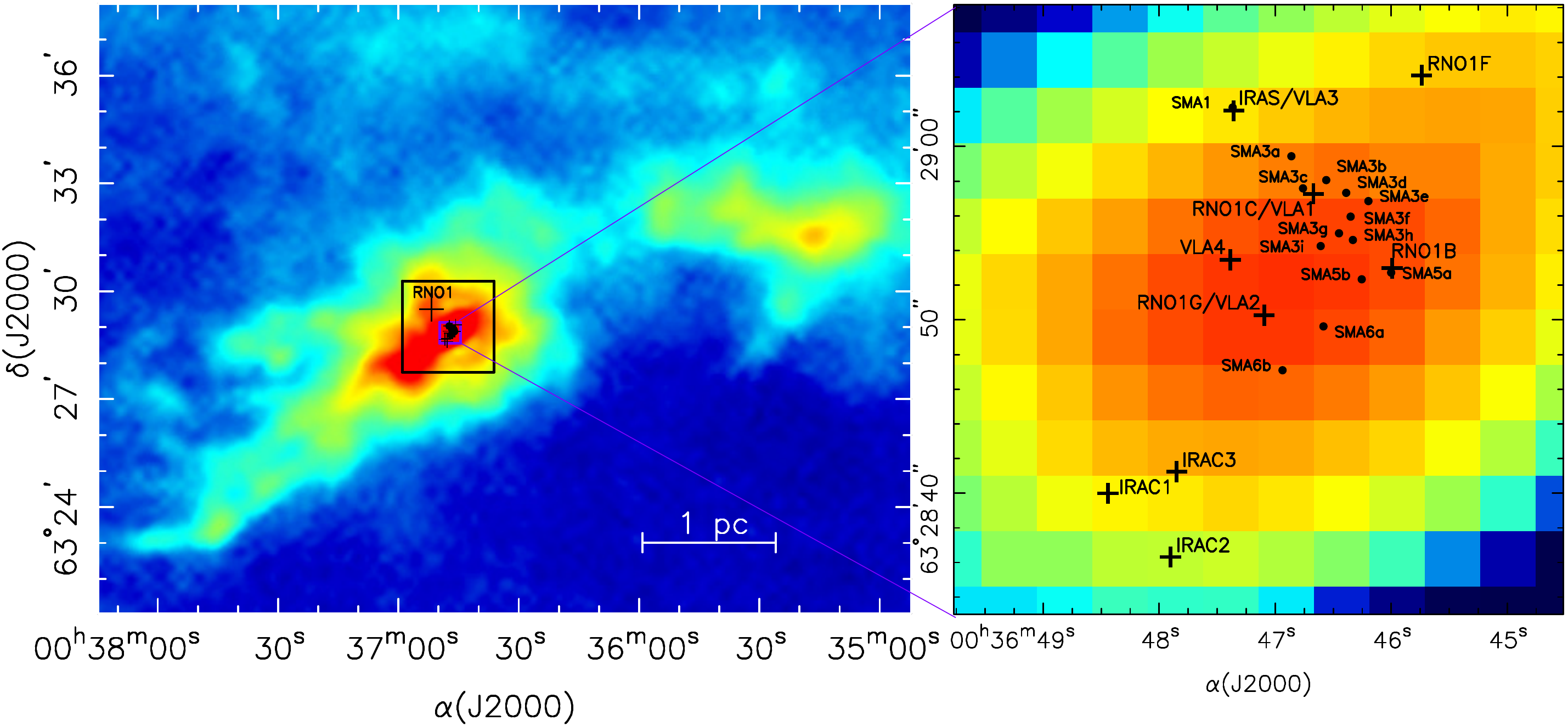}}
\caption{
\Herschel{} large scale map (left) of \hh{} column density \citep{Jua19}. 
The large-scale filament, roughly in the southeast-northwest direction, can be clearly seen.
The large box shows the region mapped in \nh{}, encompassing the area of highest column density of the filament. 
The embedded sources detected in the IR, millimeter and centimeter continuum are identified in the enlarged view (right) of the small box in the central area (except RNO 1, which lies outside the small box).
The symbols in the right panel correspond to
``$\bullet$'': millimeter dust-continuum emission \citep{Jua19}, and
``$+$'':       mid-IR sources \citep{Qua07} and centimeter continuum sources \citep{Ang94}; 
\label{fherschellog10N}}
\end{figure*}

Filaments are prevailing structures in star-forming regions as revealed by the large-scale images obtained by the \Herschel{} Space Observatory \citep[e.g.,][]{And10, Mol10, Arz11, Stutz2015,Sch20}. 
Most of the known prestellar and protostellar dense cores lie along such filamentary structures \citep[e.g.,][]{Pol13, Kon15}. 
Filaments and their intersections, the so-called hubs, are thought to constitute the essential bricks building molecular clouds \citep[e.g.,][]{Mye09}. 
In this scenario, filaments collapsing towards the densest parts of molecular clouds lead to the formation of stellar clusters \citep[e.g., ][]{Kru18}. 

\object{LDN 1287} (hereafter L1287) is the densest part of a large-scale ($\sim 100$ pc) filament traced by dust emission at 160, 250, 350, and 500 $\mu$m observed by \Herschel{} \citep[][see Fig.~\ref{fherschellog10N}]{Sch20,Jua19}, and is located at a distance of $929^{+34}_{-33}$ pc \citep{Ryg10}. 
The dense gas in the region has been mapped in
HCN, \hco{} \citep{Yan91},  
\nh{} \citep{Est93, Sep01, Sep11},  
CS \citep{Yan95, McM95}, and  
dust continuum emission \citep{San01}.
The mass in the central 0.5 pc of radius has been estimated from the C$^{18}$O emission to be $\sim 120$ \Msol{} \citep{Ume99}.

L1287 contains an energetic bipolar molecular outflow with its axis in the northeast--southwest direction \citep{Sne90, Yan91, Xu06, Feh17, Jua19}. 
A very cold and luminous IRAS point source, IRAS 00338+6312, with an estimated bolometric luminosity of 760~\Lsol{} \citep{Con07} is located towards the center of the outflow. 
A binary FU Ori system, RNO 1B/1C, whose components are separated $\sim6''$, is also found a few arcsec southwest of the nominal IRAS position \citep{Sta91, Ken93}. 
Given the large uncertainty in the IRAS position a debate was raised to clarify whether the IRAS source and the FU Ori system were the same objects, or the IRAS source actually traces an independent object \citep{Wei93}, presumably younger and more embedded than the optically visible FU Ori system \citep[classified as Class II objects;][]{Qua07}; 
and related to this, which is the driving source of the molecular outflow. 
This issue was clarified by the results of the radio continuum observations by \citet{Ang94} who detect a 3.6 cm counterpart, VLA~3, of the IRAS source. 
The accurate position of the radio source clearly indicates the presence of an embedded young stellar object independent of the FU Ori binary, a result further supported by the detection of \hho{} maser emission associated with VLA~3 \citep{Fie95, Fie97} and the more recent detection of a mid-IR counterpart \citep{Qua07}, providing a refined IR position, which is fully in agreement with the centimeter position. 
The source VLA 3, as imaged by \citet{Ang94} shows the characteristics of a thermal radio jet, and is elongated along the outflow axis, making it the best candidate to power the molecular outflow. 
This association with the outflow has been confirmed by recent SMA CO (2--1) maps of the outflow by \citet{Jua19} that show that VLA~3 is the source closest to the origin of the bipolar outflow
(see Section \ref{sec_cluster}).

Several studies propose that L1287 harbors a cluster in the process of formation. 
In addition to the protostar traced by the IRAS/VLA source and the binary FU Ori system, there are three additional VLA sources (one of them, VLA 1, very close to the position of RNO 1C) that could trace YSOs. 
More recently, \citet{Qua07} find four additional mid-IR sources tentatively classified as Class 0/I and Class II objects (one of them, the Class 0/I RNO 1G, very close to the position of VLA 2), which suggest that RNO 1B/C belong to a young, small stellar cluster, making them the only well-studied and confirmed FU Ori stars that apparently belong to a cluster-like environment. 
Interestingly, both FU Ori stars, RNO 1B and C, have been detected in X rays with \textit{Chandra} \citep{Ski20}. 
\citet{Jua19}, from 1.3 mm continuum SMA images find a large number of dust peaks, some of them close to the positions of the YSOs identified at other wavelengths, inferring a very high fragmentation level in the L1287 core (see right panel of Fig.~\ref{fherschellog10N}), as compared to fragmentation in other samples of massive dense cores. 
Also, \citet{Jua19} propose, based on both large-scale and small-scale kinematic motions of L1287, that converging flows towards the center of the filamentary structure are responsible for the formation of the cluster of YSOs, but other interpretations cannot be ruled out to explain the complex kinematics, such as outflow feedback and superposition of individual velocity components. 

In this paper we report optical H$\alpha$ and \sii{} observations, and VLA observations of the ammonia $(1,1)$ and $(2,2)$ inversion transitions with an angular resolution of $\sim3\farcs5$, to study the kinematics and physical properties of the dense gas at the center of L1287. The data presented here allowed us to disentangle at least 3 different velocity components which seem to be interacting with each other and feeding the central region of the long filament of L1287. 
In \S \ref{sec_obs} we describe the observations and the spectral line analysis,
in \S \ref{sec_results} we present the results,
in \S \ref{sec_discussion} we discuss the results obtained 
for the Guitar Core and the Blue and Red Filaments, 
and
in \S \ref{sec_conclusions} we give our conclusions.
Finally,
in Appendix \ref{achanmaps} we show additional observational data, 
in Appendix \ref{amodel} we develop the self-gravitating isothermal cylindrical filament model, used to fit the Blue and Red Filaments structure,
and
in Appendix \ref{atff} we summarize the expressions of the free-fall velocity and free-fall time for a cylindrical filament.

\section{Observations and data analysis}
\label{sec_obs}

\subsection{VLA \nh{} observations}

Observations of the $(J,K)=(1,1)$ and $(J,K)=(2,2)$ inversion transitions of the ammonia molecule (at the rest frequencies 23.694495 and 23.722633 GHz, respectively) were carried out with the VLA of the NRAO\footnote{
The National Radio Astronomy Observatory is a facility of the National Science Foundation, operated under cooperative agreement by the Associated Universities, Inc} 
in the D configuration in 1996 August 31 
(Project code: AA198; P. I.: G. Anglada).
The phase center of the interferometer was set at the position of the catalog position of IRAS 00338+6312, 
\RA{00}{36}{47}{7}, \DEC{+63}{29}{02}.

The absolute flux calibrator was 3C48 (0134+329) for which a flux density of 1.1 Jy at the observed frequency of 23.7 GHz was adopted. The phase calibrator was 0224+671, with a bootstraped flux density of 1.82 Jy, and 0316+413 was used as bandpass calibrator.

The \nh{} $(1,1)$ and $(2,2)$ lines were observed simultaneously using the four-IF spectral mode of the VLA, which allows the observation of two polarizations for each line. A bandwidth of 3.125 MHz was used, with 63 channels of 48.828 kHz width ($0.618$ \kms\ at $\lambda=1.3$ cm) centered on $V_\mathrm{LSR}=-18.0$ \kms. Calibration and data reduction were performed using standard procedures of the Astronomical Imaging Processing System (AIPS) of the NRAO. Cleaned maps were obtained with the task IMAGR of AIPS. The resulting synthesized beam size, after natural weighting of the visibility data, was $3\farcs69\times3\farcs29$ ($\mathrm{PA}=1\fdg6$) and the achieved $1\sigma$ noise level per spectral channel was 3 \mjyb (equivalent to 0.54 K of beam-averaged brightness temperature).

\subsection{CAHA Optical observations}

The L1287 field was observed with the 2.2~m telescope of the Calar Alto Observatory (CAHA) in 2016 November, within a program aimed to image optical counterparts of IRAS-sources YSOs candidates 
\citep[Autumn 2016 \#12; P.I.: A. Riera; ][]{Lop20}.

Deep CCD narrow-band images were obtained using the Calar Alto Faint Object Spectrograph (CAFOS) with the SITe chip, which gives an image scale of $0\farcs53$~pixel$^{-1}$ and a field of view of $\sim~16'$. Three narrow-band filters were used: 
an \halpha\ filter (central wavelength $\lambda=6569$~\AA, bandpass $\Delta\lambda=50$ \AA), 
an \sii\ filter (central wavelength $\lambda=6744$~\AA, bandpass $\Delta\lambda=97$ \AA) that included the emission from the \sii{} $\lambda=6717, 6731$ \AA{} lines, and 
an additional filter of central wavelength $\lambda=6607$ \AA, bandpass $\Delta\lambda=43$ \AA) to obtain the continuum adjacent to the \halpha{} and \sii{} emission lines.

We obtained images in \halpha{} and \sii{} of 1 hr of total integration time by combining three frames of 1200 s exposure each and an additional continuum image of 1200 s integration after combining two frames of 600 s exposure.
All the images were processed with the standard tasks of the IRAF\footnote{
IRAF is distributed by the National Optical Astronomy Observatories, which are operated by the Association of Universities for Research in Astronomy, Inc., under cooperative agreement with the National Science Foundation.}
reduction package, which included bias subtraction and flatfielding corrections using sky flats. In order to correct for the misalignment among individual frames, the frames were recentered using the reference positions of field stars well distributed around the source. 

Astrometric calibration of the images was performed in order to compare the optical emission with the positions of the radio continuum sources. The images were registered using the $(\alpha, \delta)$ coordinates from the USNO Catalogue\footnote{
The USNOFS Image and Catalogue Archive is operated by the United States Naval
Observatory, Flagstaff Station.} 
of ten field stars well distributed on the observed field. The rms of the transformation was $0\farcs2$ in both coordinates. 

\subsection{NH$_3$ Hyperfine Fitting analysis}
\label{sec_hfs}

The \nh{} $(1,1)$ and $(2,2)$ lines were analyzed by means of the Hyperfine
Structure (HfS) tool (Estalella 2017). This tool fits simultaneously the
hyperfine quadrupole and magnetic structure of the \nh{} $(1,1)$ and $(2,2)$ inversion
transitions, with the assumptions that the beam filling factor, the excitation 
temperature, the hyperfine line width, and the central velocity are the
same for all the hyperfine lines. 
In addition, HfS uses the results of the fit to derive physical parameters
including the excitation temperature, \nh{} column  density, rotational and
kinetic temperature, with the assumption that the emitting region is homogeneous
along the line of sight.

\begin{figure}[htb]
\resizebox{\hsize}{!}{\includegraphics{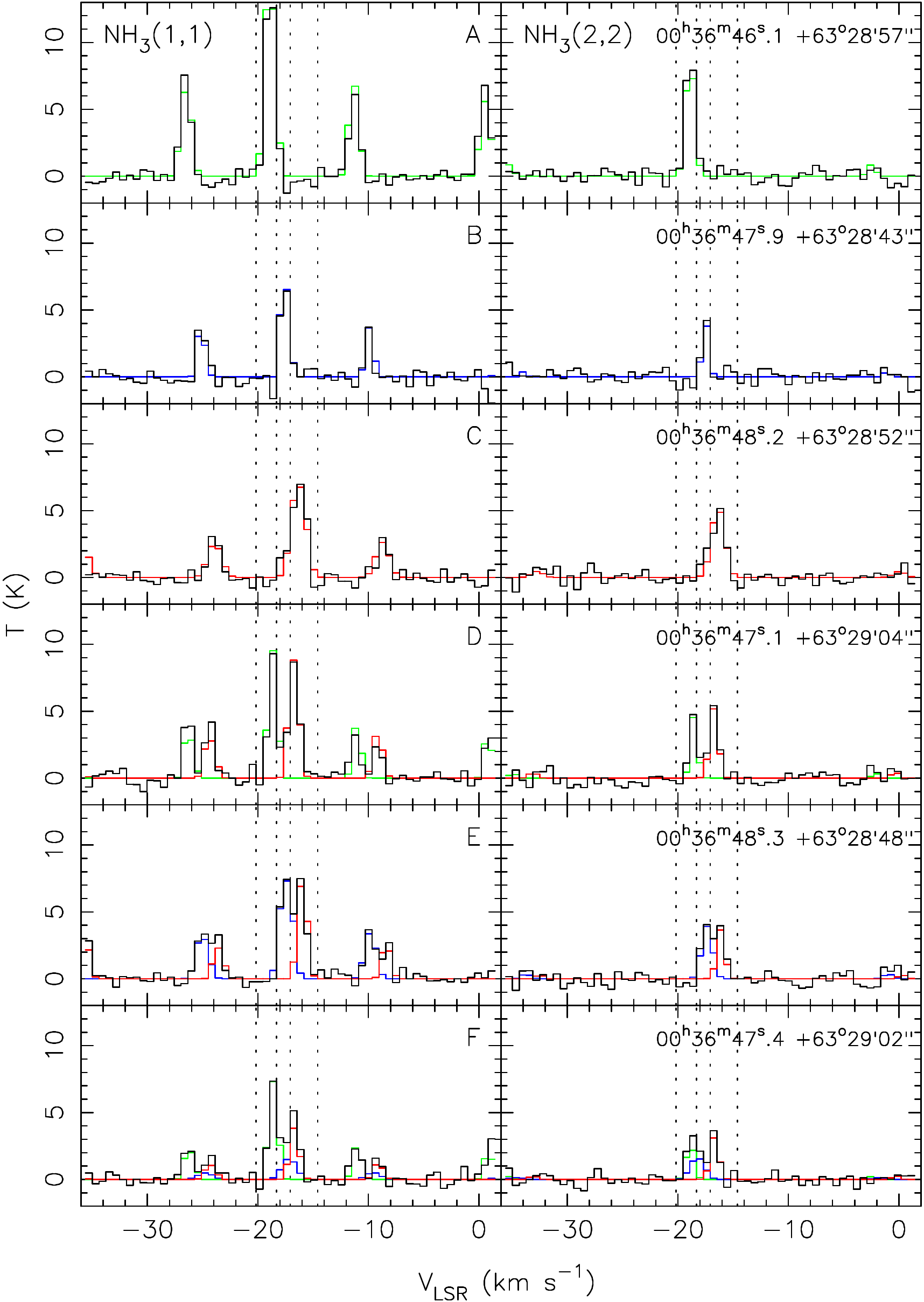}}
\caption{
Examples of \nh{} $(1,1)$ (left) and $(2,2)$ (right) spectra in L1287: observed (black), fitted component 1 (green), component 2 (blue), component 3 (red). The dotted vertical lines show the range of velocities used for fitting the three velocity components. From top to bottom: 
position A, component 1 only;
B, component 2 only;
C, component 3 only;
D, components 1 and 3;
E, components 2 and 3;
F, all three components present.
The coordinates of the six positions are given in the right panels.
\label{hfs_spectra}}
\end{figure}

Despite the relatively low spectral resolution of the observations (0.618 \kms), the spectra showed the presence of several velocity components in the emission, which were apparent in different positions of the region (see also the channel maps in Figs.\ \ref{fchanmap11}, \ref{fchanmap11sat}, and \ref{fchanmap22}). 
After careful inspection of the spectra, three different velocity components were identified, with non-overlapping velocity ranges, valid for all the region mapped. 
The range of velocities for each component were 
$-20.16$ to $-18.31$ \kms\ (component 1),
$-18.31$ to $-17.07$ \kms\ (component 2), and
$-17.07$ to $-14.60$ \kms\ (component 3).
In Fig.\ \ref{hfs_spectra} we show the $(1,1)$ and $(2,2)$ spectra at selected positions where only one of the components is present (three top rows) or two or all three components are present (three bottom rows). 
The positions where the spectra were taken are given in the figure.

The HfS tool was used to fit simultaneously three velocity components of the  $(1,1)$ and $(2,2)$ transitions, each one inside the velocity range given above. 
The fit was performed for the central 256$\times$256 pixels of the map.
For each pixel (pixel size $0\farcs6$), the $(1,1)$ and $(2,2)$ spectra were averaged within a boxcar of $3''$ in diameter. 
We checked that no information was lost with the smoothing, and the fits were more reliable due to the enhanced signal-to-noise ratio.
The fit was performed for pixels where the main beam brightness temperature of both lines was higher than 2.0 K (4 times the typical rms noise level). Some examples of the fits for the different components are shown in Fig.\ \ref{hfs_spectra}.
In total, the fit was performed for 3728 pixels, with 
component 1 fitted in 1318 pixels, 
component 2 in 2537 pixels, and 
component 3 in 1005 pixels.
For each component fitted in each pixel, six line parameters were obtained from the fit,
1: hyperfine-line full-width at half-maximum (deconvolved from the channel width), $\Delta V$, the same for the $(1,1)$ and $(2,2)$ transitions;
2: $(1,1)$ central velocity, $V_\mathrm{LSR1}$;
3: amplitude $A$\footnote{
The amplitude is defined as $A=f[J_\nu(\Tex)-J_\nu(\Tbg)]$, where $f$ is the beam filling factor, 
$J_\nu(T)=(h\nu/k)/[\exp(h\nu/kT)-1]$ is the Planck correction to the Rayleigh-Jeans law, 
$\Tex$ is the excitation temperature, and 
$\Tbg$ is the background temperature.
} 
times the $(1,1)$ main component optical depth, $A\tau_{1m}$;
4: $(1,1)$ main component optical depth, $\tau_{1m}$;
5: $(2,2)$ central velocity, $V_\mathrm{LSR2}$; and
6: amplitude $A$ times the $(2,2)$ main component optical depth, $A\tau_{2m}$.
In addition, physical parameters like \T{rot}, \T{k}, $N(\mathrm{NH_3})$ were obtained
\citep[see][for a detailed description of the parameters]{Est17},
which were used in the analysis of the physical characteristics of the region.
FITS images for the fitted parameters for each velocity component were produced by HfS.

With the aim of characterizing the kinematics of the gas, an additional run of the HfS three-component fit was performed to fit only the $(1,1)$ spectra, irrespective of the intensity of the $(2,2)$ emission. 
For this run, the fit was performed for 8420 pixels, with
component 1 fitted in 2262 pixels,
component 2 in 6633 pixels, and
component 3 in 2858 pixels.
For pixels fitted in both runs ($(1,1)$ and $(2,2)$ simultaneously, and $(1,1)$ only) the values of the $(1,1)$ line parameters fitted in the second run showed no significant differences from the values fitted in the first run.

The spectra with best signal-to-noise ratio have fit uncertainties 
in $V_\mathrm{LSR}$ of the order of 0.02--0.11 \kms{},
in $\Delta V$ of the order of 0.05--0.09 \kms{}, and
in \T{rot} of the order of 1.4--3.8 K.
The errors were greater for spectra with not so good signal-to-noise ratio, and the median values of the errors in $V_\mathrm{LSR}$, $\Delta V$, and \T{rot} are given in Table \ref{tparameters}.
The results were inspected for bad fits. As a general rule, fits with values of $\Delta V > 1.5$ \kms, or errors in  $\Delta V > 1.0$ \kms, which in general corresponded to fits of a low-intensity, very wide component, were considered an artifact and blanked out.
After blanking, 3593 pixels remained with $(1,1)$ and $(2,2)$ fits, and 8371 pixels with fits of the $(1,1)$ line only.

\section{Results}
\label{sec_results}

The morphology of the region is dominated by a large-scale filament, several parsec long, in a roughly southeast-north west direction, clearly visible in the \hh{} column density (Fig. \ref{fherschellog10N}) and dust temperature maps obtained from \Herschel{} data \citep{Jua19}. 
Our VLA ammonia observations show details of the dense gas distributed in extended filaments in a region of $\sim1\farcm5$ in diameter encompassing both RNO 1, a background young star with 2MASS counterpart, at a distance of $1.24\pm0.05$ kpc \citep{Gaia18}, and the RNO 1B/C objects.
The \halpha{} and \sii{} images show strong emission associated with the RNO 1 and RNO 1B/C objects.

\subsection{\halpha{} and \sii{} emission}

\begin{figure}[t]
\resizebox{\hsize}{!}{\includegraphics{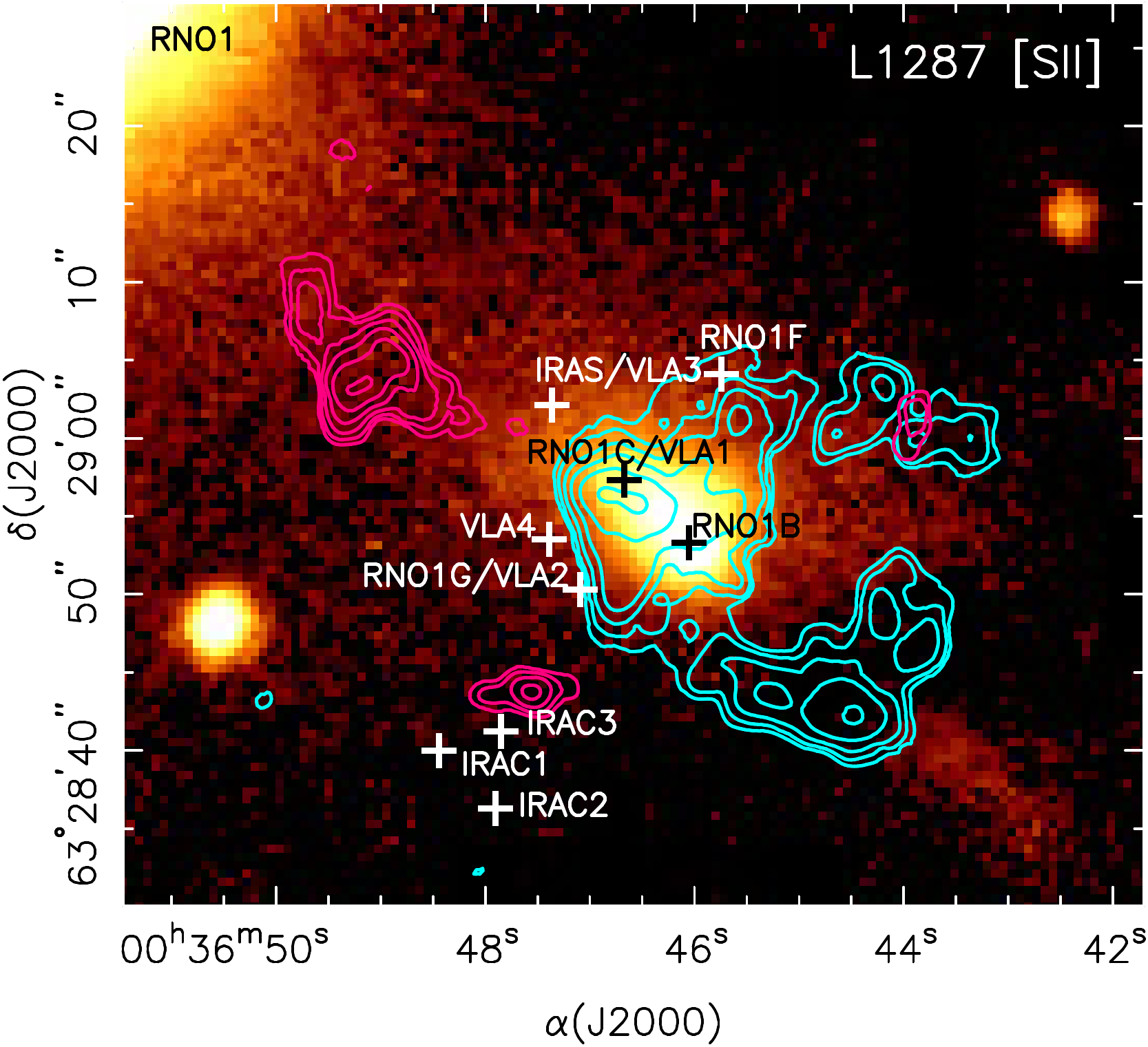}}
\resizebox{\hsize}{!}{\includegraphics{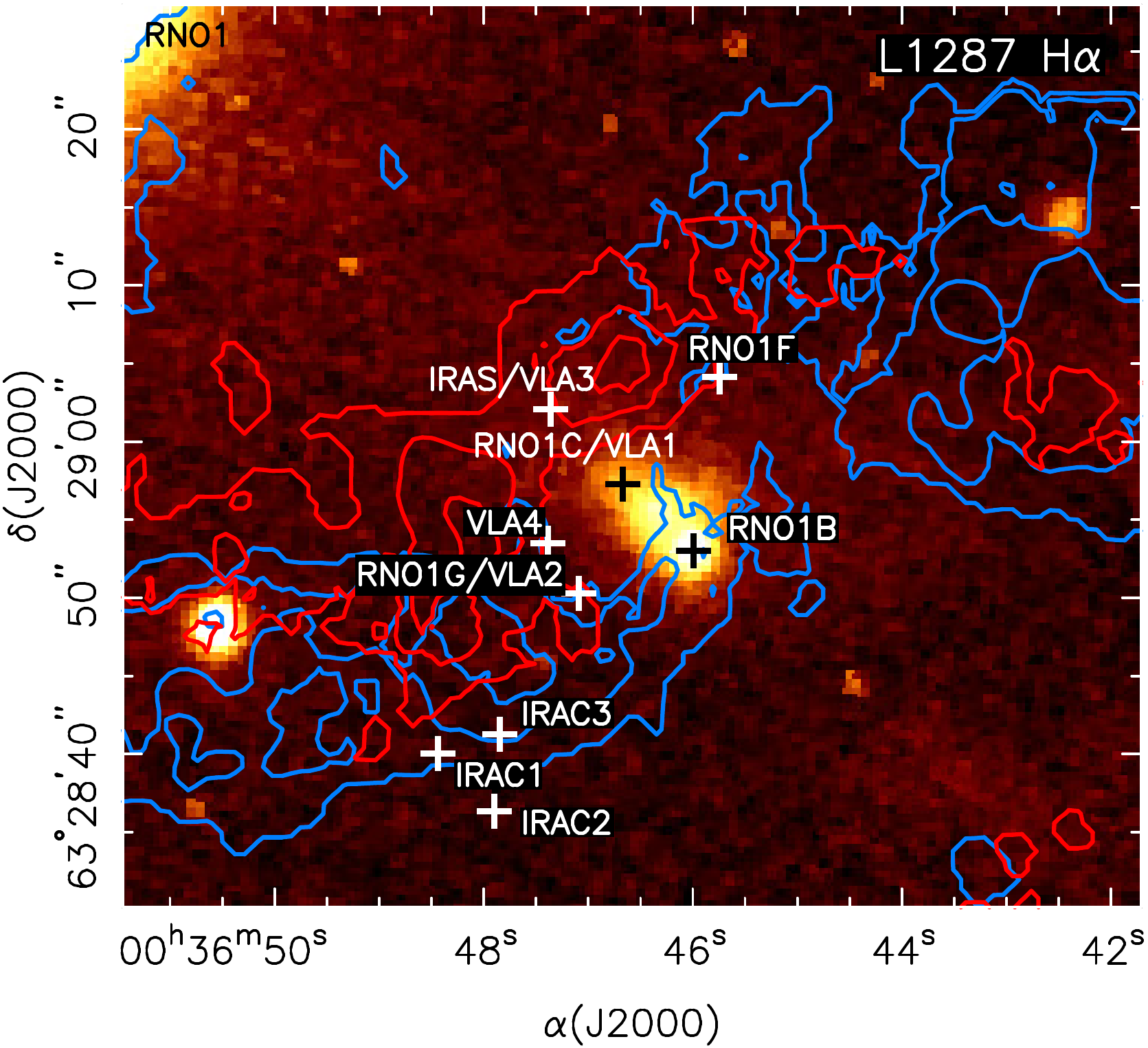}}
\caption{
Images of L1287 obtained with the CAHA 2.2 m telescope. 
The IR and centimeter continuum sources are identifies and their positions are identified with the black or white crosses. 
\emph{Top:} \sii{} (color scale);
the blue and red contours trace the small-scale CO outflow observed by \citet{Jua19}.
\emph{Bottom:} \halpha{} (color scale); 
contours trace the integrated line emission of \nh{} $(1,1)$ from the Blue and Red Filaments (this work).
\label{caha}}
\end{figure}

The \halpha{} and \sii{} images obtained show a very similar morphology. 
The image of the \sii{} emission, characteristic of shock-excited ionization, is presented in Fig.\ \ref{caha} 
in the top panel and that of \halpha{} in the bottom panel.
The strong emission in the upper-left corner corresponds to RNO 1, and two field stars appear near the left and right borders of the image. 

Regarding the emission at the center of the image, the strongest \sii{} 
and \halpha{} 
emission comes from a compact ($\sim10''$ or $\sim9\,000$ au) region, which encompasses the positions of the two stars of the FU Ori binary RNO 1B/C, while the positions of IRAS/VLA 3, RNO 1F, RNO 1G/VLA 2, and VLA 4 fall in the periphery, where the emission is weaker.
The optical emission at the central region is most probably tracing the central cavity detected in the near-IR by \citet{Ken93}, where the extinction is lower.

There is faint \sii{} 
and \halpha{}
emission elongated in the direction of the axis of the bipolar CO outflow, and appears associated with its blue lobe as mapped by \citet{Jua19} (see Fig.\ \ref{caha}). 
There is no emission detected in association with the outflow red lobe, due probably to a higher extinction.

\subsection{Ammonia emission: characterization of the different velocity components}

\begin{figure}[htb]
\resizebox{\hsize}{!}{\includegraphics{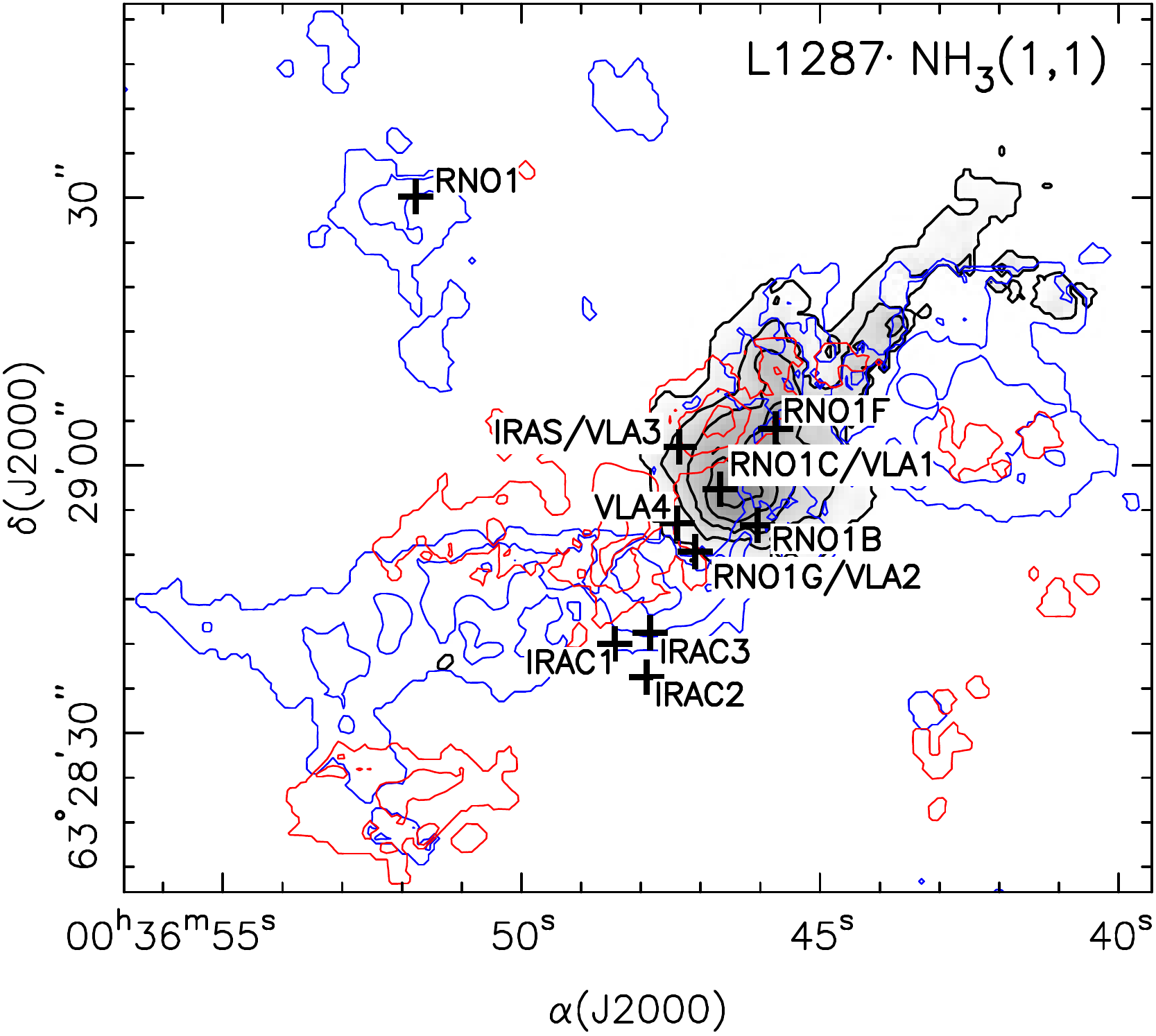}}
\caption{
\nh{} $(1,1)$ integrated intensity map of the three velocity components in L1287. In order of increasing velocity, 
black contours and gray scale}: component 1 (Guitar Core), $-20.16$ to $-18.31$ \kms; 
blue contours: component 2 (RNO 1 and Blue Filament), $-18.31$ to $-17.07$ \kms; 
red contours: component 3 (Red Filament), $-17.07$ to $-14.60$ \kms.
\label{hfs_rgb}
\end{figure}

\begin{table*}[htb]
\small
\centering
\caption{Guitar and filaments parameters
\label{tparameters}}
\begin{tabular}{lccccccccccrccc}
\hline\hline
&
&
&
\multicolumn{2}{c}{$V_\mathrm{LSR}$\tablefootmark{a}} &
\multicolumn{2}{c}{$\Delta V$\tablefootmark{a}} &
\multicolumn{2}{c}{\T{rot}\tablefootmark{a}} &
\multicolumn{2}{c}{$N(\mathrm{NH_3})$\tablefootmark{a}} &
\\
Core /&
Velocity range &
Size &
\multicolumn{2}{c}{(\kms)} & 
\multicolumn{2}{c}{(\kms)} & 
\multicolumn{2}{c}{(K)} &
\multicolumn{2}{c}{($10^{15}$ \cmd)} &
$M$\tablefootmark{b} & 
\T{k}\tablefootmark{c} &
$c_s$\tablefootmark{d} & 
\\ 
Filam.& 
(\kms) &
(pc) &
Med. & 1 $\sigma$& 
Med. & 1 $\sigma$& 
Med. & 1 $\sigma$& 
Med. & 1 $\sigma$& 
(\Msol) & 
(K) &
(\kms) & 
$\cal{M}$\tablefootmark{d} 
\\ 
\hline
RNO 1      & $-18.31,-17.07$ & 0.07$\times0.07$ & $-17.53$ & $0.13$ & $0.27$	& $0.15$ & $15.4$ & $1.2$ & $1.37$ & $0.26$ & $\sim10$  & 16.9 & 0.26 & 0.50 \\
Guitar     & $-20.16,-18.31$ & 0.26$\times0.12$ & $-18.60$ & $0.18$ & $0.41$ & $0.22$ & $16.7$ & $2.2$ & $1.32$ & $0.76$ & 30  & 18.7 & 0.26 & 0.97 \\
Blue Fil.  & $-18.31,-17.07$ & 0.52$\times0.10$ & $-17.50$ & $0.31$ & $0.30$ & $0.20$ & $17.0$ & $4.3$ & $1.11$ & $0.48$ & 104 & 19.1 & 0.26 & 0.56 \\
Red Fil.   & $-17.07,-14.60$ & 0.40$\times0.10$ & $-16.75$ & $0.41$ & $0.40$ & $0.29$ & $20.2$ & $4.1$ & $0.78$ & $0.50$ & 13  & 23.9 & 0.29 & 0.77 \\
\hline
\end{tabular}
\tablefoot{ 
\tablefoottext{a}
{Median value and 1-$\sigma$ equivalent dispersion (half-difference between the 84 and 16 percentiles) for all the pixels with a fitted spectrum for each core or filament.}
\tablefoottext{b}
{Obtained by integration of the \nh{} column density maps after applying the primary beam correction, except for RNO~1 where the mass was taken as the product of the average column density and the area with \nh{} $(1,1)$ emission. The ammonia abundance  assumed was $X(\mathrm{NH_3})=1\times10^{-8}$.}
\tablefoottext{c}
{Kinetic temperature, obtained from \T{rot} \citep{Est17}.}
\tablefoottext{d}
{Sound speed and Mach number (see text).}
}
\end{table*}

In Appendix \ref{achanmaps} we show the channel maps of the observed \nh{} $(1,1)$ main line (Fig.\ \ref{fchanmap11}) and inner satellites (Fig.\ \ref{fchanmap11sat}), and the \nh{} $(2,2)$ line (Fig.\ \ref{fchanmap22}). 
The positions of the known YSOs proposed to be associated with the region are indicated in these figures. 
The \nh{} $(1,1)$ and $(2,2)$ spectra towards the positions of these YSOs are shown in Fig.\ \ref{fspectra_vla}. 

In Fig.\ \ref{hfs_rgb} we show the integrated intensity map of the three velocity components of the \nh{} emission identified in L1287 through the HfS analysis. 
In order of increasing velocity, 
component 1 (hereafter ``Guitar'' Core, see discussion below) is shown in gray scale and black contours, 
component 2 (corresponds to the RNO 1 core and the ``Blue Filament'') in blue contours, and 
component 3 (hereafter ``Red Filament'') in red contours.

In general, the three velocity components trace different parts of the high-density gas, with a few locations where there is superposition of more than one component (see Figs.\ \ref{hfs_spectra} and \ref{hfs_rgb}).
There are pixels in these regions for which the assignment of emission to the 2nd or 3rd velocity components could not be done accurately with the present spectral resolution. 
This occurs mainly in
i) the southwards extension of the eastern part of the Blue Filament, where there is emission of the southeast blob of the Red Filament, and 
ii) the westernmost part of the Blue Filament, where there is also some emission from the Red Filament (see Fig.{} \ref{hfs_rgb}).

In Table \ref{tparameters} we give the median values and 1-$\sigma$ equivalent dispersion for all the pixels with a fitted spectrum for RNO 1, the Guitar Core, and the Blue and Red Filaments, of the values of $V_\mathrm{LSR}$, $\Delta V$, \T{rot}, and $N(\mathrm{NH_3)}$. 

\begin{figure}[htb]
\centering
\resizebox{0.9\hsize}{!}{\includegraphics{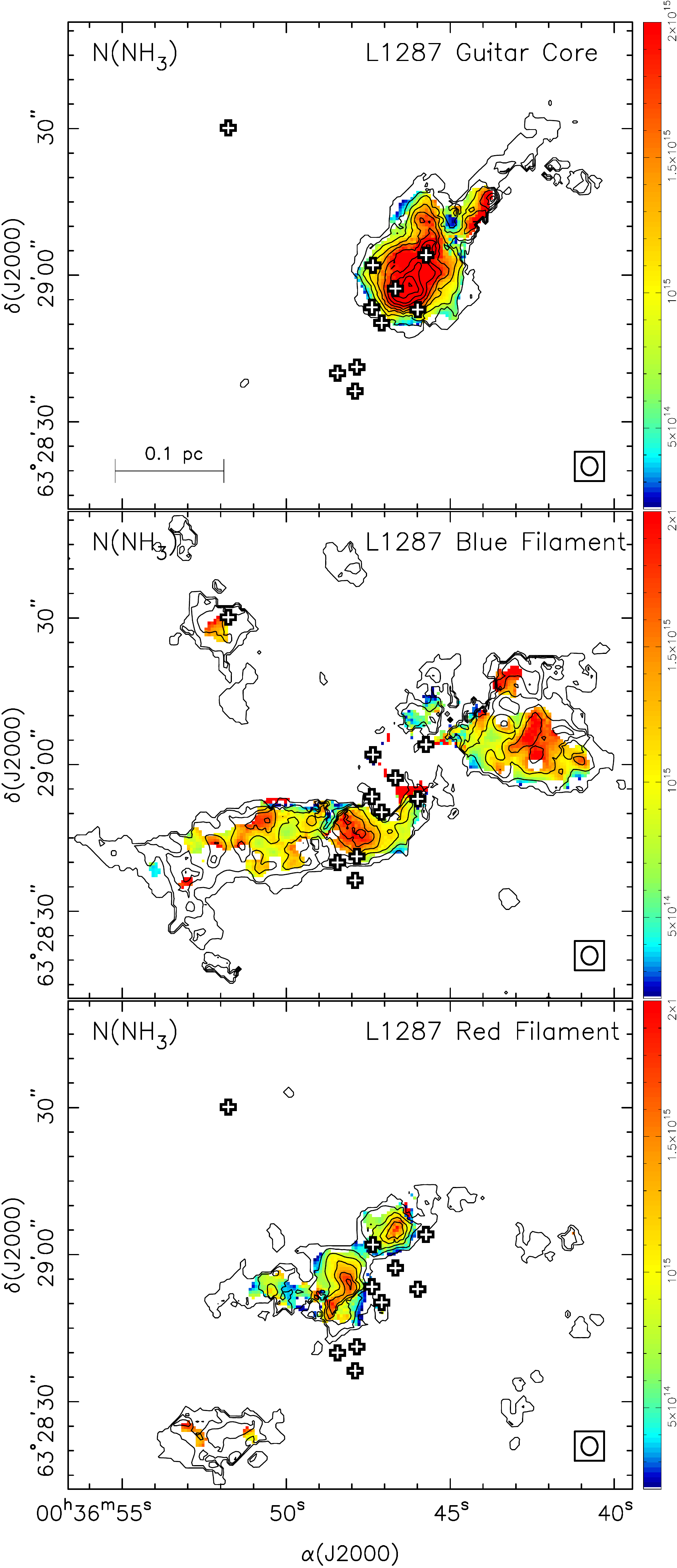}}
\caption{
L1287 \nh{} beam-averaged column density, corrected for primary beam response, derived from \nh{} $(1,1)$ and $(2,2)$, assuming a beam filling factor $f=1$ (color scale).
The color scale ranges from $2\times10^{14}$ cm$^{-2}$ (blue) to $2\times10^{15}$ cm$^{-2}$ (red).
The contours are the \nh{} $(1,1)$ integrated line-intensity, corrected for opacity, $A\tau_m\Delta V$, of the
Guitar Core (top, velocity range $-20.16$ to $-18.31$ \kms), 
RNO 1 and Blue Filament (middle, $-18.31$ to $-17.07$ \kms), and 
Red Filament (bottom, $-17.07$ to $-14.60$ \kms).
Contours are from 1 to 37 in steps of 4 K~\kms.
The crosses correspond to YSOs identified in Fig.{} \ref{caha}.
\label{hfs_nnh3}
\label{hfs_ataumdv}
}
\end{figure}

The \nh{} $(1,1)$ integrated line-intensity maps corrected for opacity ($A\tau_m\Delta V$) of the three components are shown in Fig.\ \ref{hfs_ataumdv} in contours. 
The embedded YSOs are also indicated in this figure.
As can be seen in the figure, the emission from the Guitar Core (top panel) is compact (the body of the guitar), with a diameter of $\sim26''$ (0.12 pc) plus an extension to the northwest (the neck of the guitar), with a length of $\sim36''$ (0.16 pc).
The emission of the Blue and Red Filaments (middle and bottom panels), on the contrary, is elongated and twisted. 
The length of the Blue filament is of the order of $120''$ (0.52 pc), and that of the Red Filament of $90''$ (0.40 pc). The width of both filaments is of the order of $20''$ (0.10 pc).
The length and width of the Blue and Red filaments in \nh{} are much smaller than that of the dusty filamentary structure, as recently reported in the \Herschel{} HiGAL catalogue \citep{Sch20}.
The Herschel dust filament has a length of 6.1~pc and a width of 0.24~pc. Such larger values are expected since the dust filament seen by Herschel extends well beyond the region mapped in NH3, which traces the densest part (close to the spine) of the filament.
The main part of the Blue Filament (middle panel) is roughly in the east-west direction, and bends northwards at the location of the embedded sources.
A gap in the \nh{} emission, at the position of the embedded YSOs, separates this main part from another clump of emission with no clear elongation, to the west of the main part, in the direction of the large-scale filament traced by \Herschel{} (see Fig.\ \ref{fherschellog10N}). 
The Red Filament (bottom panel) has a general direction southeast to northwest, compatible with the filament oriented at $\mbox{PA}=34\degr4$ reported in the Hi-GAL catalogue \citep{Sch20}. 

Fig.\ \ref{hfs_nnh3} also shows, in color scale, the beam-averaged \nh{} column density, calculated assuming a beam filling factor $f=1$. 
In this and following figures, the portion of the figures which shows contours but no color scale are pixels which were fitted with only the \nh{} $(1,1)$ line.
The median value for all pixels fitted for each component, and the 1-$\sigma$ dispersion are given in Table \ref{tparameters}. 
As can be seen, the median column density of the Guitar Core is higher than that of the filaments.
The masses of the Guitar and filaments have been obtained from the integration of the column density maps, after applying the primary beam correction (see Table \ref{tparameters}). 
The values obtained assuming an ammonia abundance $X(\mathrm{NH_3})=1\times10^{-8}$ are 
30  \Msol{} (Guitar Core), 
104 \Msol{} (Blue Filament), and 
13  \Msol{}\ (Red Filament). 

\begin{figure}[htb]
\centering
\resizebox{0.9\hsize}{!}{\includegraphics{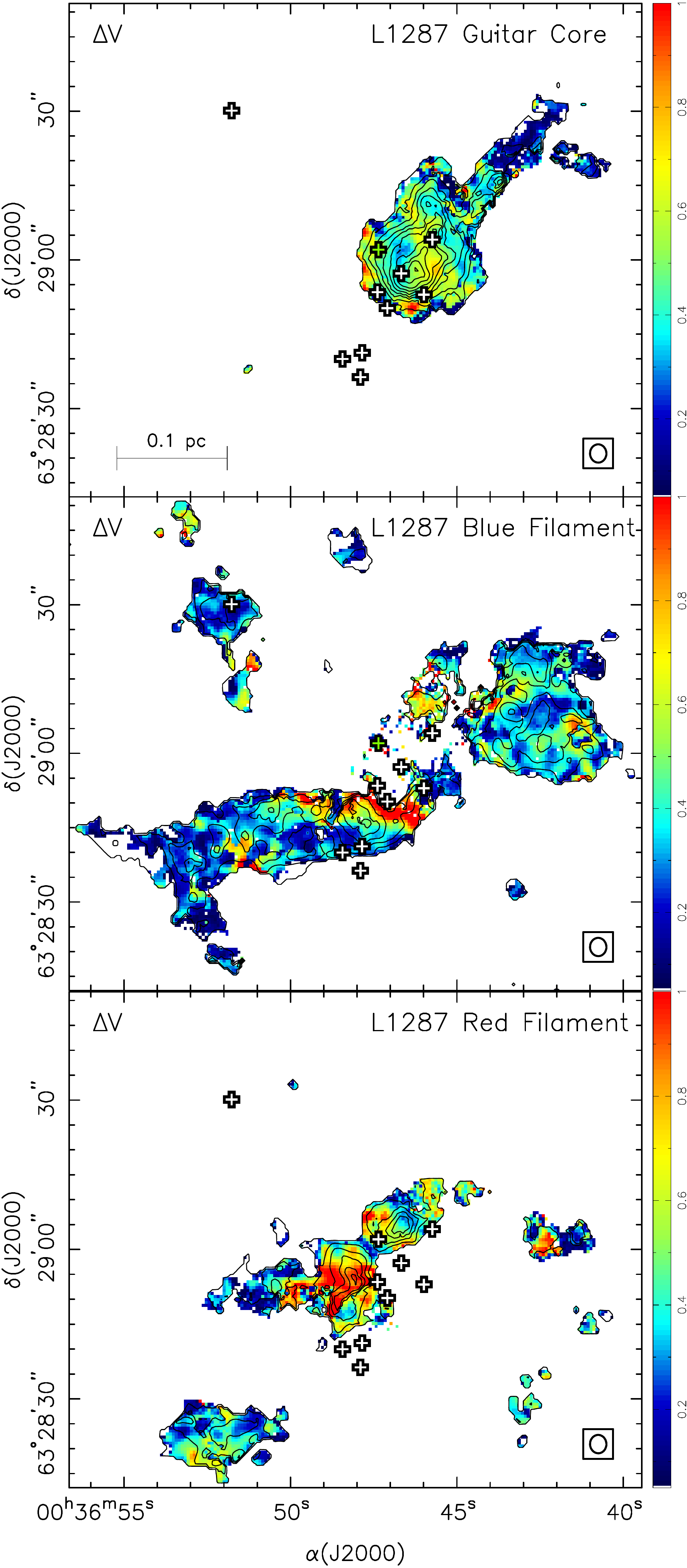}}
\caption{
L1287 \nh{} hyperfine-line full-width at half-maximum, $\Delta V$ (color scale), 
The color scale ranges from 0.1 \kms{} (blue) to 1.0 \kms{} (red). 
The contours are the same as in Fig.\ \ref{hfs_ataumdv}.
The crosses correspond to YSOs identified in Fig.\ \ref{caha}.
\label{hfs_dvline}
}
\end{figure}

\begin{figure}[htb]
\centering
\resizebox{0.9\hsize}{!}{\includegraphics{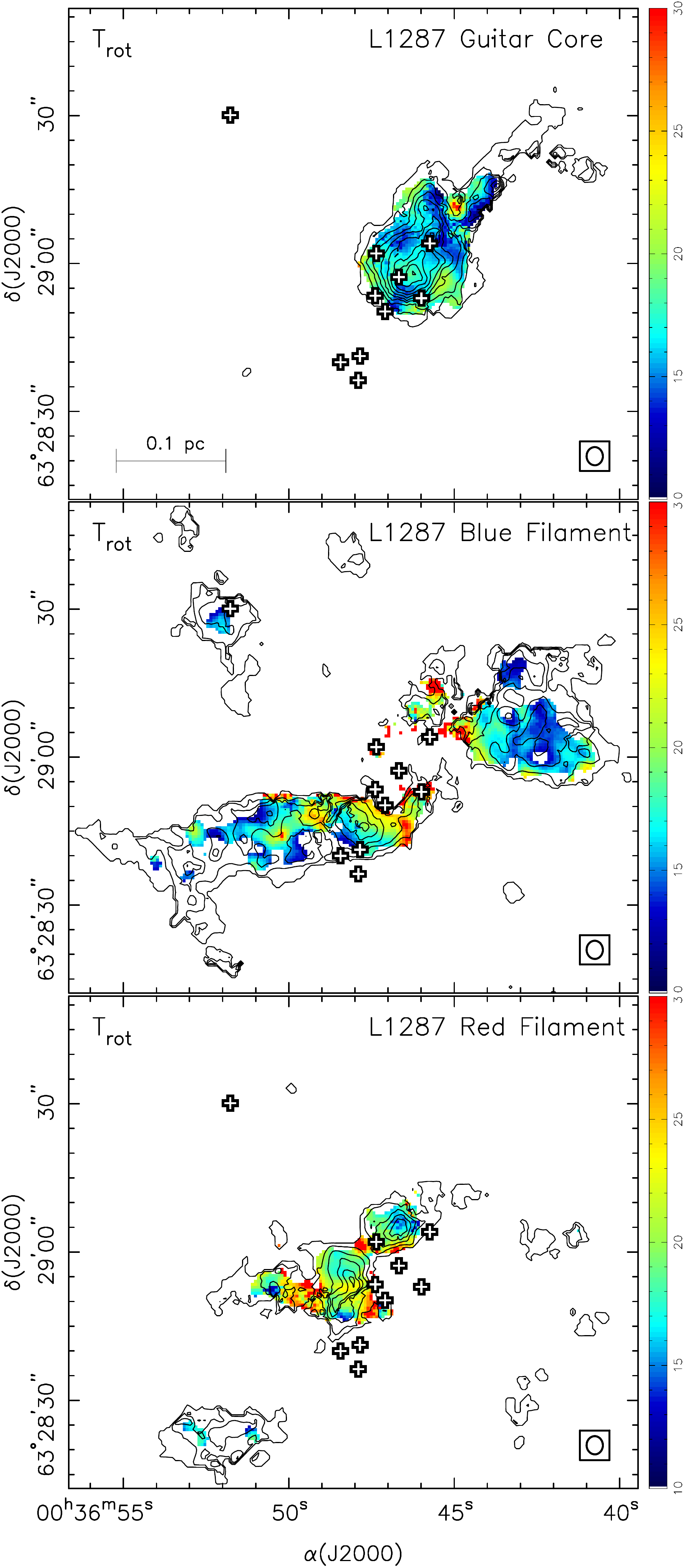}}
\caption{
L1287 rotational temperature, \Trot{}, derived from \nh{} $(1,1)$ and $(2,2)$ (color scale), 
The color scale ranges from 10 K (blue) to 30 K (red). 
The contours are the same as in Fig.\ \ref{hfs_ataumdv}.
The crosses correspond to YSOs identified in Fig.\ \ref{caha}.
\label{hfs_trot}
}
\end{figure}

The velocity component from $-18.31$ to $-17.07$ \kms{} also traces emission from a core encompassing the position of RNO~1, $\sim40''$ ($\sim0.17$ pc) north of the main part of the Blue Filament (middle panel in Figs.\ \ref{hfs_nnh3}, \ref{hfs_dvline}, and \ref{hfs_trot}). 
The RNO 1 core presents narrow lines ($\Delta V \simeq 0.27$ \kms{}), with a central velocity of $-17.5$ \kms{}, and a low rotational temperature ($\sim15$ K). 
Regarding the mass of RNO 1, since the \nh{} $(2,2)$ emission was detected in only a few pixels, the mass was estimated as the product of the average column density of RNO 1 (corrected for primary beam response) and the area with \nh{} $(1,1)$ emission, resulting in a value of $\sim10$ \Msol{} (see Table \ref{tparameters}),

\begin{figure}[htb]
\centering
\resizebox{0.9\hsize}{!}{\includegraphics{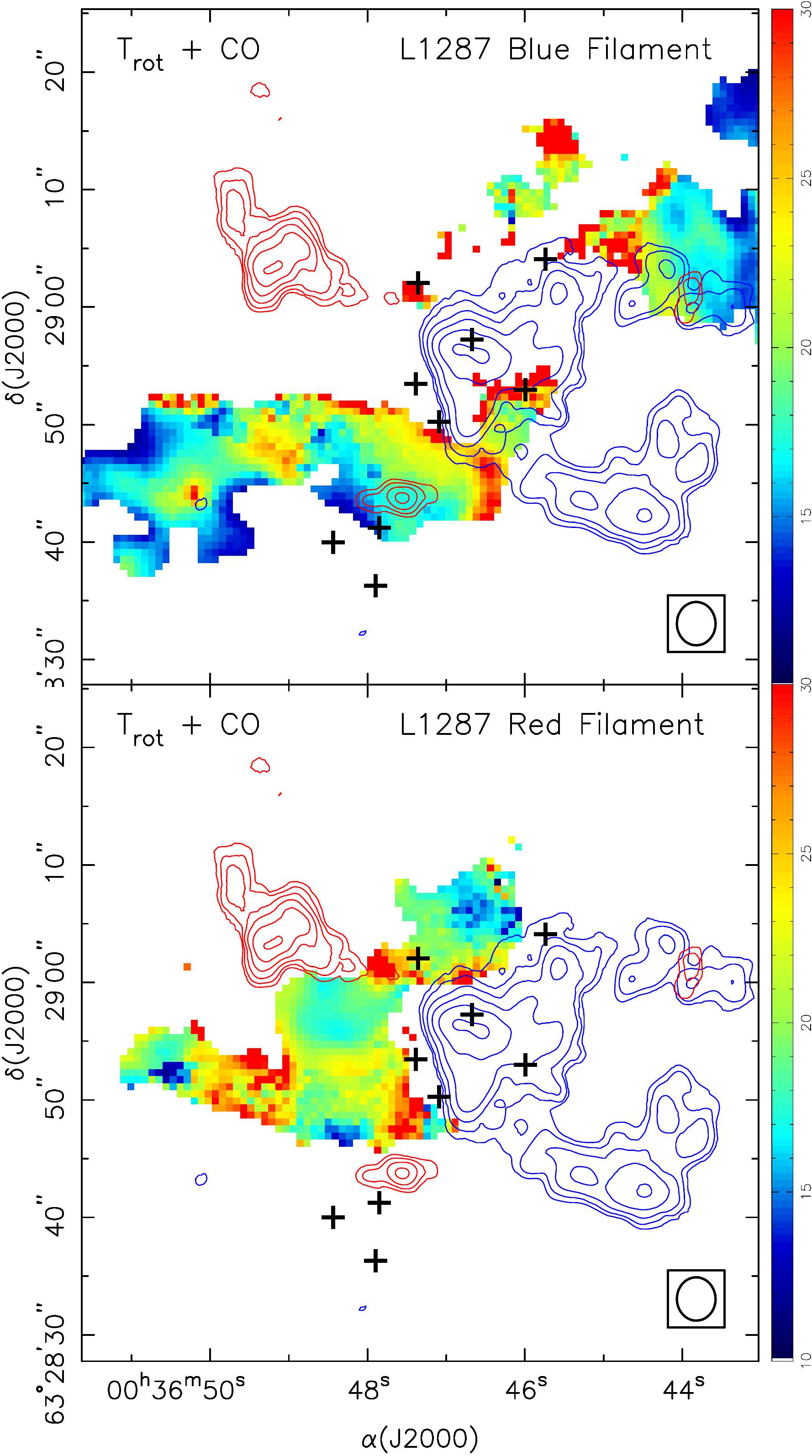}}
\caption{
Zoom of the L1287 rotational temperature of the Blue Filament (top) and Red Filament (bottom), superimposed to the high-velocity CO emission (blue and red contours).
There is a clear temperature increase at the walls of the cavity where the blue lobe of the outflow propagates.
The crosses are the same as in Fig.\ \ref{caha}.
\label{hfs_trot_zoom}
}
\end{figure}

In Fig.\ \ref{hfs_dvline} we show the maps of hyperfine-line full-width at half-maximum, $\Delta V$, obtained from the \nh{} $(1,1)$ fits, with the same color scale, for the Guitar Core (top) and the Blue and Red Filaments (middle and bottom). 
The median values of the line width are given in Table \ref{tparameters}.
In all the \nh{} structures analyzed the hyperfine lines have narrow linewidths. 
It is remarkable that the individual kinematic features in the channel maps of Figs.{} \ref{fchanmap11}, \ref{fchanmap11sat}, and  \ref{fchanmap22} are present only for one or sometimes two spectral channels.

In Fig.\ \ref{hfs_trot} we show the maps of the rotational temperature, $T_\mathrm{rot}$, obtained from the \nh{} $(1,1)$ and $(2,2)$ fits, with the same color scale for the Guitar Core (top) and the Blue and Red Filaments (middle and bottom). 
The difference between the Guitar and the filaments is remarkable. 
While for the Guitar the rotational temperature is almost constant, with a median of 16.7 K with an extremely low 1-$\sigma$ dispersion of 2.2 K, for the filaments the median rotational temperature is 17--20 K, with a 1-$\sigma$ dispersion significantly higher, $\sim 4$ K (see Table \ref{tparameters}). 
There are pixels in both filaments, near the positions of the embedded sources, with rotational temperatures in excess of 40~K. 
This can be clearly seen in the zoom of the rotational temperature map near the embedded sources of Fig.\ \ref{hfs_trot_zoom}.
However, \Trot{} becomes highly uncertain for values higher than $\sim30$ K using only $(1,1)$ and $(2,2)$ data due to a lack of information on higher energy transitions \citep{Fri17}.

\section{Discussion}
\label{sec_discussion}

\subsection{The cluster of embedded YSOs near RNO 1B/C}
\label{sec_cluster}

A small cluster of YSOs has been identified in the central part of L1287 from optical, IR and radio observations. 
However, some uncertainties regarding the identification of the counterparts at different wavelengths, and the number of distinct objects, still remain. 
The VLA observations of \citet{Ang94} provide source positions at 3.6 cm with an accuracy of $\sim0.2''$ and the observations of \citet{Qua07} provide mid-IR positions with an accuracy of $\sim0.3''$. 
These two sets of observations unambiguously show that the young embedded object traced by IRAS 00338+6312 coincides with VLA 3 and is a different object than the FU Ori binary RNO 1B/C. 
On the other hand, \citet{Ang94} tentatively associate the radio source VLA 1 with the FU Ori star RNO 1C. 
However, at that time the optical position of the star was known with an accuracy of only $\sim2''$ \citep{Sta91, Wei93}, making uncertain the association. 
Now, after Gaia DR2 \citep{Gaia18} the optical positions of the two FU Ori stars, RNO 1B and 1C, are known with a precision of a fraction of mas, revealing that the positions of VLA 1 and RNO 1C differ by $\sim0.4''$. 
Despite the positions are now much better constrained, the difference between radio and optical positions is only slightly greater than the accuracy of the VLA position ($0.2''$), and thus we consider that the association remains still uncertain. 
The radio source VLA 2 falls within $0.4''$ of the position of the mid-IR source RNO 1G \citep{Qua07}, so their positions are marginally consistent within the uncertainties. 
More accurate radio and IR positions are necessary for an unambiguous identification of these counterparts.
The radio source VLA 4 has no known IR counterpart, but it seems to be associated with dust and compact molecular emission \citep{Jua19}. 

Therefore, there are between 9 and 11 known independent objects 
(the Class 0/I objects
IRAS/VLA 3, 
RNO 1G/VLA 2, 
IRAC 1, 
IRAC 2;
the Class II objects
RNO 1B, 
RNO 1C/VLA 1, 
RNO 1F, 
IRAC 3; 
and
VLA 4),
of which at least 8 are very likely YSOs, within a region of $\sim30''$ (30\,000 au = 0.15 pc in projection) towards the central part of L1287.

The \nh{} channel maps (Figs.\ \ref{fchanmap11}, \ref{fchanmap11sat}, and \ref{fchanmap22}) show a clumpy distribution of the emission, suggesting that different dense gas clumps, spatially and kinematically separated from each other, could be associated with the different YSOs of the proposed still-forming cluster \citep[e.g.,][]{Qua07}. 
The clumpy nature of the \nh{} emission and possible association with the central young cluster is better seen in the $-18$ \kms{} channel map of the \nh{} $(1,1)$ main line shown in Fig.\ \ref{fchanmap11}, where a one-to-one association of ammonia clumps with IRAS/VLA 3, RNO 1B, and RNO 1C/VLA 1 is suggested. 
It is worth noting that the positions of the three ammonia clumps coincide with three of the dust cores recently reported by \citet{Jua19} from 1.3 mm SMA continuum observations. 
Actually, \citet{Jua19} identify in the central region of L1287 a total of 14 dust cores (some of them associated with molecular line emission), with masses of a fraction of solar mass, in their $1''$ resolution SMA continuum observations \citep[][see right panel of Fig.~\ref{fherschellog10N}]{Jua19}, inferring a very high degree of fragmentation in this central region. 
We also note that RNO 1F and IRAC 3 appear close to ammonia peaks in the $-18.6$ and $-17.4$ \kms{} channel maps, respectively, but their association with this dense gas is less clear. 

This specific association between dust/molecular cores and YSOs is less clear in the images obtained after separating the \nh{} emission in three components. 
Some of the dust emission peaks might be just column density enhancements resulting from the superposition of several independent features along the line of sight instead of representing real density enhancements in cores. 
It is also possible that some cores remain unidentified in \nh{} because of the difficulties in fitting the different velocity components given the sensitivity, angular resolution, and spectral resolution of the present \nh{} observations.
We note that the positions of the sources IRAS/VLA 3, RNO 1B, RNO 1C/VLA 1, and RNO 1F fall towards the densest part of the Guitar Core, and that the positions of RNO 1G/VLA 2 and VLA 4 fall near its south-east edge. 
However, there are no detectable signs of local heating or perturbation in the line width and rotational temperature maps towards these positions (Figs.\ \ref{hfs_dvline} and \ref{hfs_trot}). 
Therefore, we conclude that, unless the luminosity of these objects is insufficient to produce a detectable perturbation, they are not associated with the Guitar Core, which is likely observed in projection. 

The role of IRAS/VLA3, the likely more embedded YSO and best candidate to drive the outflow, is particularly relevant. 
This source appears directly associated with \nh{} emission from the Red Filament, as shown in the bottom panel of Fig.\ \ref{hfs_ataumdv}, and with signs of dynamical perturbation, traced by an increased line width (Fig.\ \ref{hfs_dvline}), and local heating (Figs.\ \ref{hfs_trot} and \ref{hfs_trot_zoom}). 
However, because of the low spectral resolution and blending of line components the interaction of IRAS/VLA 3 with dense gas remains still uncertain. 
New observations with higher angular and spectral resolution are necessary to clarify the role of this object with respect to the filaments.

The IR/radio sources appear to be distributed near the edges of the filaments. 
In particular, there is an apparent cavity, with a diameter of $\sim12\,000$ au towards the central region of the Blue/Red Filaments, associated with \halpha{} and \sii{} emission.
RNO 1C/VLA 1 is located at its center, and the sources IRAS/VLA 3, RNO 1B, RNO 1G/VLA2, VLA 4, and RNO 1F are located near its inner edge (see middle and lower panels in Figs.\ \ref{hfs_ataumdv}, \ref{hfs_dvline}, and \ref{hfs_trot}, as well as the $-17.4$ to $-16.6$ \kms{} channel maps of Figs.\ \ref{fchanmap11}, \ref{fchanmap11sat}, and \ref{fchanmap22}). 
The presence of such a cavity in the L1287 core was proposed by \citet{Ken93} from near-IR data , and by \citet{Yan95} and \citet{McM95} from CS data. 
Our \halpha{}, \sii{}, and VLA ammonia observations provide a confirmation with data of higher quality.  
There are signs of dynamical perturbation and local heating, revealed by an increase in the \nh{} line width and temperature in the Blue Filament near the positions of RNO 1G/VLA 2 and VLA 4 (middle panel in Fig.\ \ref{hfs_dvline}). 
RNO 1B appears projected towards the \nh{} emission of the Blue Filament, with no clear signs of dynamical perturbation (Fig.\ \ref{hfs_dvline}) but with a hint of local heating (Figs.\ \ref{hfs_trot} and \ref{hfs_trot_zoom}). 
The shape of the high-velocity CO outflow fits very well with the morphology of the cavity traced by the ammonia emission (Fig.\ \ref{hfs_trot_zoom}), strongly supporting that this cavity has been created by the interaction of the outflow with the dense gas of the filaments (see further discussion in Sect.\ \ref{sec_BRF}).

\subsection{The Guitar Core}
\label{sec_guitar}

As already discussed, the Guitar Core is probably seen in projection at the position of the embedded sources, and could be prestellar in nature. 
However, as will be shown in the following, there is a clear signpost of infall onto a central protostar, and thus the Guitar Core is not pre-stellar, but most probably a very young protostellar core. 

\begin{figure}[htb]
\resizebox{\hsize}{!}{\includegraphics{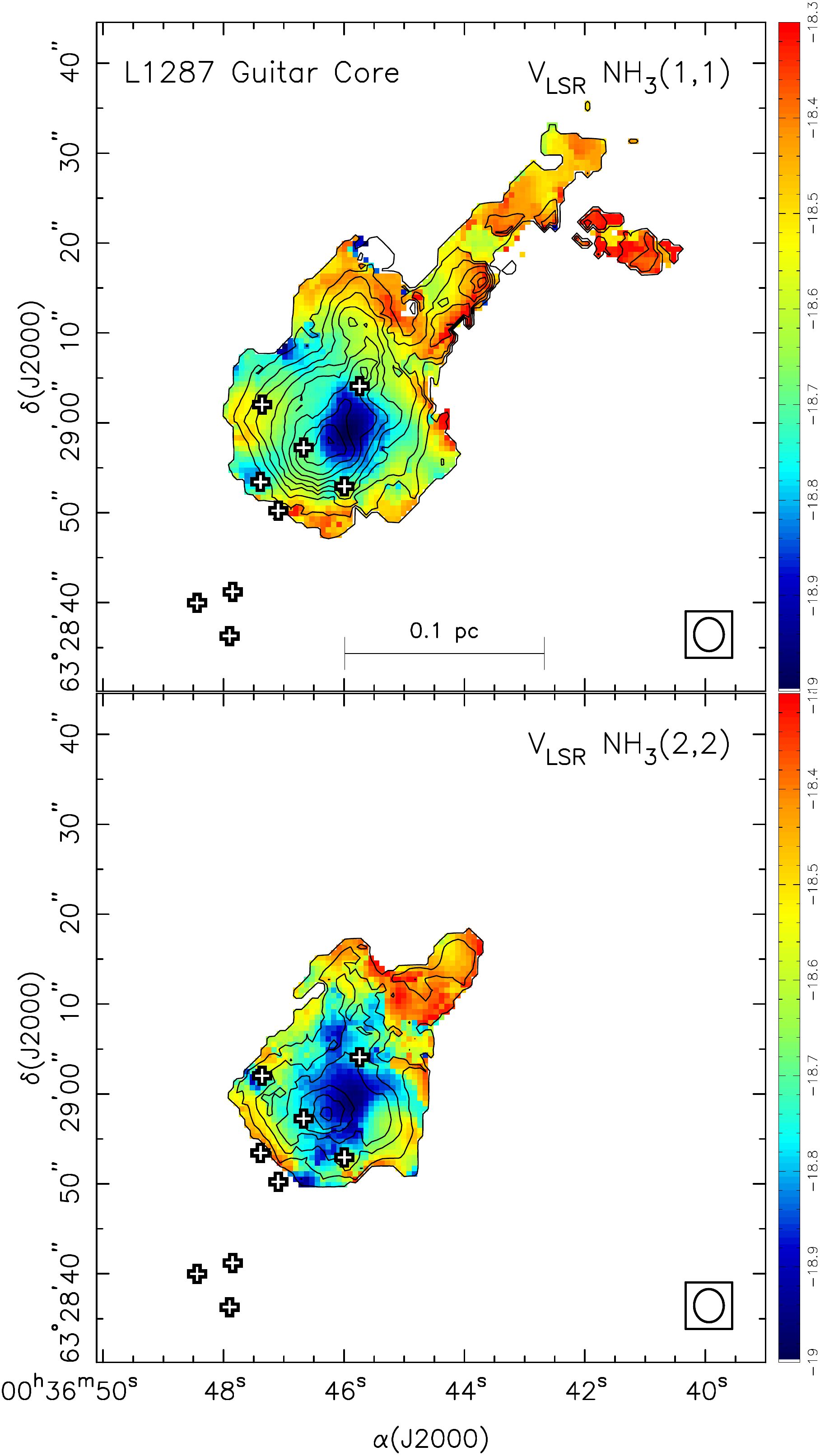}}
\caption{
Central velocity of \nh{} $(1,1)$ (top) and $(2,2)$ (bottom) lines of the Guitar Core in L1287 (color scale). 
The contours are the $(1,1)$ (top) and $(2,2)$ (bottom) integrated line emission.
The velocity at the position of the emission peak is blueshifted $\sim0.4$ km~s$^{-1}$ with respect to the rest of the gas of core. 
This blue spot at the emission peak of the core (the sound hole of the guitar) is the hallmark of infalling material (see text).
The crosses are YSOs identified in Fig.\ \ref{caha}.
\label{hfs_vlsr}}
\end{figure}

\begin{figure}[htb]
\resizebox{\hsize}{!}{\includegraphics{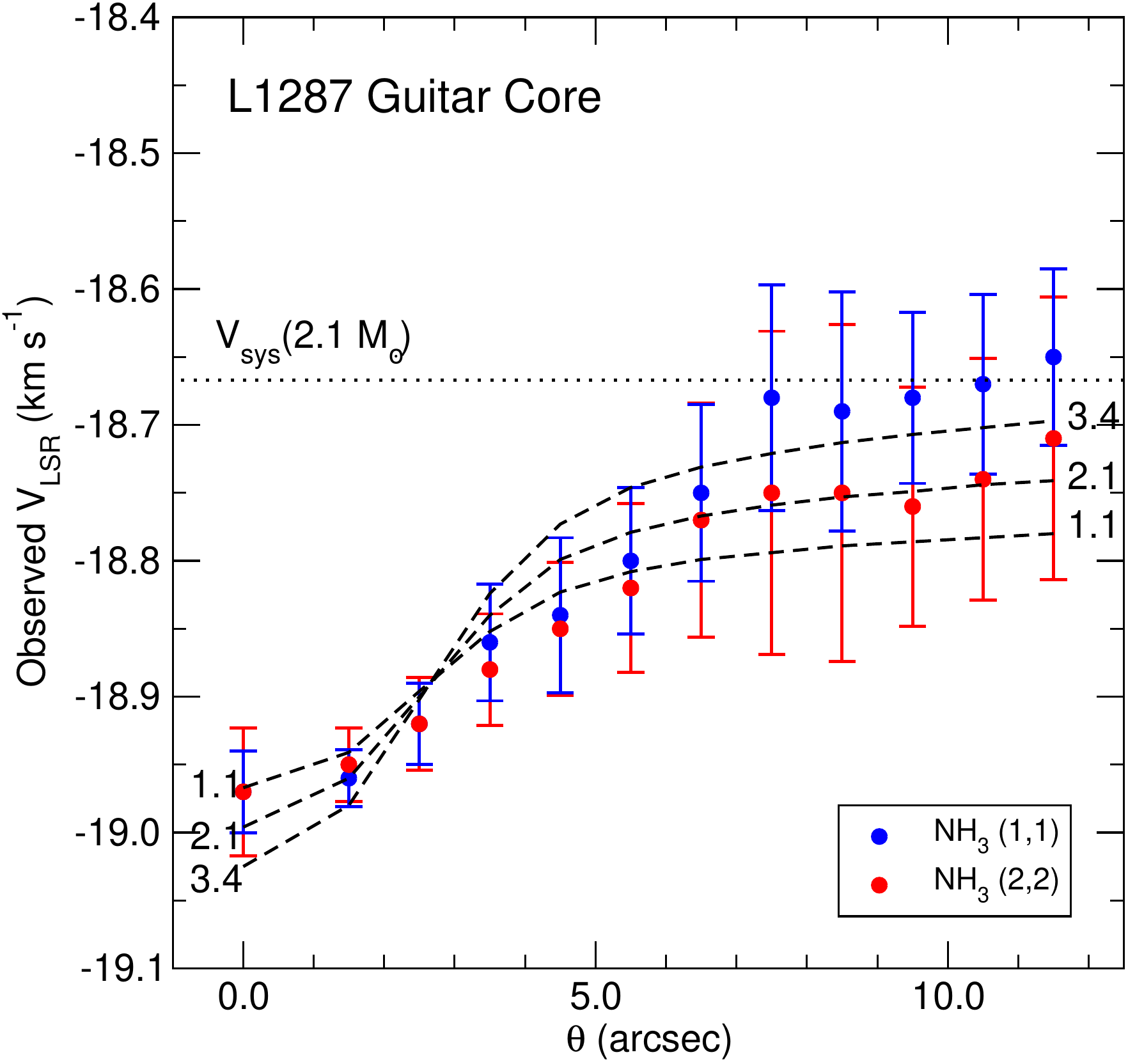}}
\caption{
Ring-averaged central velocity of the \nh{} $(1,1)$ (blue circles) and $(2,2)$ (red circles) lines in the Guitar Core in L1287. 
The error bars are the rms of the values inside each ring.
The central dashed line is the best fit of the infall hallmark model 
for a central mass of 2.1 \Msol{}, and the horizontal dotted line is the best fit systemic velocity, $V_\mathrm{sys}=-18.67$~\kms{}. The fits for central masses of 1.1 and 3.4 \Msol{} are also shown.
\label{guitar_rings}}
\end{figure}

The central velocity maps of the Guitar Core (Fig.\ \ref{hfs_vlsr}) show a spot of blueshifted gas at the center of the core (the sound hole of the guitar). 
This blueshifted gas is not produced by another clump of gas, with a blueshifted velocity, seen in projection at the center of the Guitar Core, because the line width and rotational temperature of the gas of the Guitar Core do not change 
significantly towards the blue spot (see Figs.{} \ref{hfs_dvline} and \ref{hfs_trot}).
The variation in central velocity is smooth, with no abrupt changes, as can be seen in Figs.{} \ref{hfs_vlsr} and \ref{guitar_rings}.

The blue spot can be interpreted as infalling gas toward the center of the clump  \citep{Ang91, May14, Est19}. 
The model worked out by \citet{Ang87, Ang91} and \citet{Est19} assumes an infalling core with temperature and infall velocity that are power laws of the radial distance to the protostar, with a power-law index $-1/2$. 
In addition the model assumes that the emission at any line-of-sight velocity is optically thick enough that the observer sees only the emission of the part of the isovelocity surface facing the observer. 
This produces an asymmetry in the emission, the emission from blueshifted channels being stronger than the corresponding redshifted channels. 

This asymmetry produces a negative value of the first-order moment (intensity-averaged velocity) of the emission line, which can be calculated analytically for the case of an infinite angular resolution \citep{Est19}. 
In order to model the moment 1 observed with a finite angular resolution, the radial profile has to be convolved numerically with the beam of the observation. 
The model of \citet{Est19} can also take into account the  finite radius of the collapsing core.
The model was used to fit the first-order moment maps of the \nh{} $(3,3)$, $(4,4)$, $(5,5)$, and $(6,6)$ emission mapped with the VLA in the in the G31.41+0.31 hot molecular core, 
the H$^{13}$CO$^+$ ($J=1$--0) emission observed with Nobeyama and the $^{13}$CO ($J=2$--1) emission observed with ALMA in the B335 core, 
and a preliminary analysis of the \nh{} emission in L1287 presented here,
deriving values of the central masses onto which the infall is taking place \citep{Est19}.

In L1287 the $(1,1)$ and $(2,2)$ main-component optical-depths at the position of the blue spot were $\sim4$ and $\sim2$ respectively (see Fig.\ \ref{ftau12}), and therefore high enough to produce an asymmetry between the blueshifted and redshifted channels. 
The central velocity of the \nh{} $(1,1)$ and $(2,2)$ lines for the Guitar Core, obtained from the HfS fits, was averaged in rings $1''$ wide, centered on the emission peak, up to a 
radius of $12''$.
The velocity profile obtained is shown in Fig.\ \ref{guitar_rings}. The error bars are the rms dispersion of the velocities averaged in each ring. 
As can be seen in the figure, the velocity at the origin is blueshifted $\sim 0.3$ \kms{} with respect to the ambient gas velocity. 
The model of \citet{Est19} predicts that such a velocity shift, observed with a beam of $3\farcs48$ at a distance of 929 pc, can be produced by a 
central mass of $\sim2$~\Msol.
We fitted the infalling model with an infall radius much larger than the equivalent size of the beam (3\,200 au). The best-fit values for the central mass and systemic velocity were 
$M= 2.1$~\Msol{},\footnote{
In the paper \citet{Est19}, because of a bug in the plotting program used, the ring-averaged velocities were fitted up to radial distances of $20''$, well outside of the central part of the core. 
The correct central mass for L1287 is the value given in the present paper. 
The correction does not affect values of the central mass for other cores in \citet{Est19}.
}
$V_\mathrm{sys}= -18.67$~\kms{}.
The quality of the fit is indicated by the value of the $\chi^2$ statistic for $\nu=22$ degrees of freedom (the total number of rings used in the fit, minus 2), 
$\chi^2= 11.4$, 
which gives a reduced 
$\chi_r= (\chi^2/\nu)^{1/2}=0.70$.
We checked the parameter space and found that 
$\chi_r < 0.9$ for a range of masses, 
$1.1 < M < 3.4$~\Msol{}, 
and of systemic velocities,
$-18.81 < V_\mathrm{sys} < -18.53$~\kms{}.

The conclusion is that the observed blue spot in the velocity map of the Guitar Core is the hallmark of infalling motions onto a central object, and that a simple model to interpret the data predicts that a central mass of $\sim2$ \Msol{} can explain the observed infalling motion.

\subsection{The Blue and Red Filaments}
\label{sec_BRF}

The Blue and Red Filaments show clear signs of interaction with the embedded
YSOs. In the region near the position of the sources the low
velocity resolution of the data did not allow us to separate well the two
velocity components, but significant line width and rotational temperature
enhancements are seen for both filaments. 

\begin{figure}[htb]
\centering
\resizebox{\hsize}{!}{\includegraphics{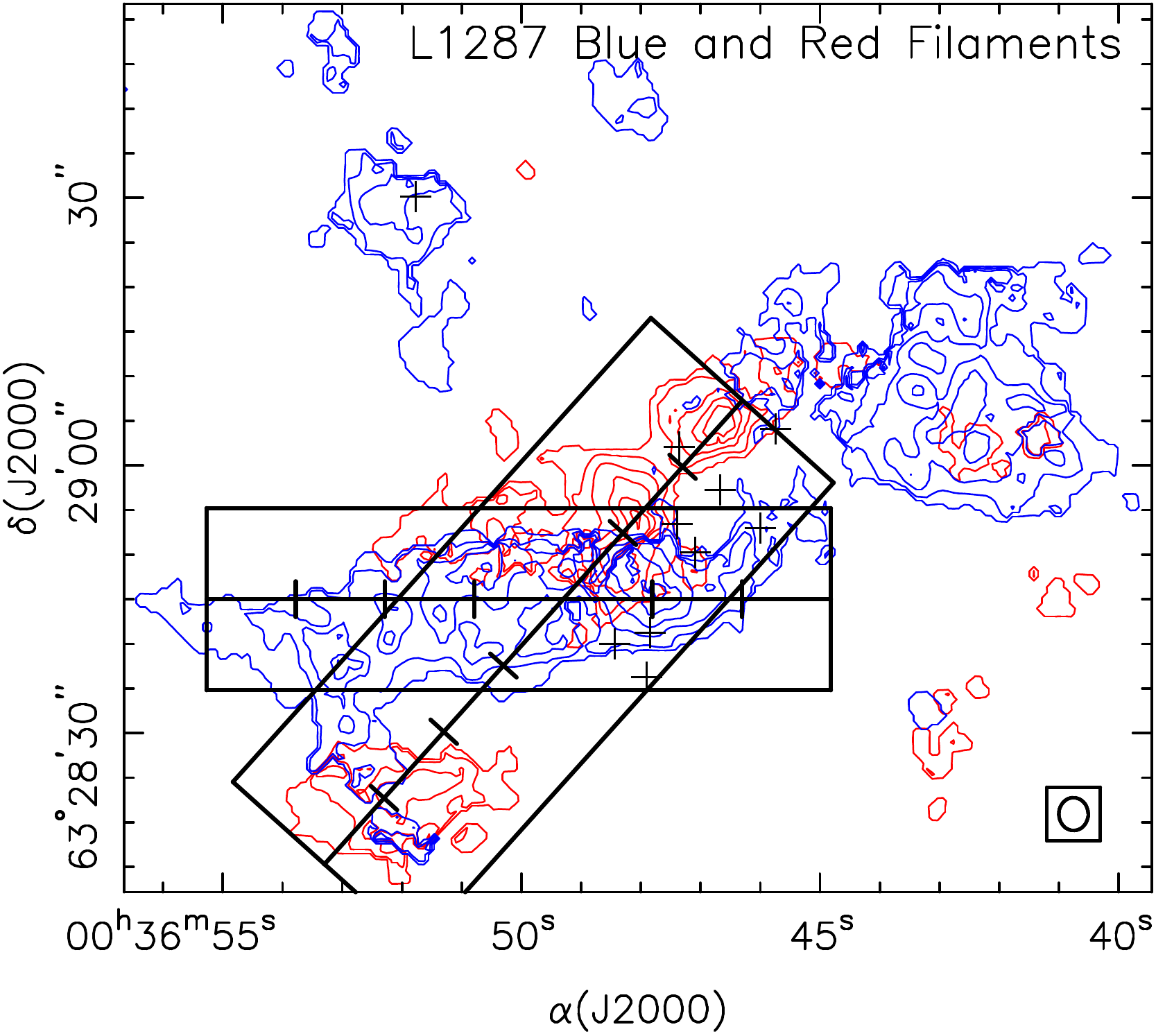}}
\caption{
Slices used to analyze the filaments, superimposed on the integrated line-intensity, corrected for opacity, $A\tau_m\Delta V$, of the Blue Filament (blue contours) and the Red Filament (red contours).
The origin of offsets along the filaments is taken at the position of the intersection of the two slices, and increase towards the west (Blue Filament) and the north-west, (Red Filament), from $-40''$ to $+30''$. Ticks in the central line of the slices are at intervals of $10''$. 
\label{fslices}}
\end{figure}

The interaction of the filaments with the star formation activity is clearly seen in Fig.\ \ref{hfs_trot_zoom}, where we show in detail the rotational temperature in the region of the filaments around the embedded YSOs, superimposed to a map of the outflow traced by high velocity CO (2--1) observed by \citet{Jua19}. 
Both filaments trace different parts of a cavity where the YSOs are located, and where the outflow propagates.
The walls of this cavity is where the highest values of the gas temperature are found, indicating that the gas heating is probably due to the interaction of the high-velocity gas with the high-density ambient gas traced by the \nh{} emission, as seen in other outflow cavity walls \citep[e.g., IRAS 20293+3952,][]{Pal07}.

Although both filaments are bent, there are sections almost straight and easy for making a slice to analyze the structure of the filaments.
The slice along the Blue Filament was at a position angle $\mathrm{PA}=-90\degr$, and that along the Red Filament at $\mathrm{PA}=-42\degr$ (see Fig. \ref{fslices}). 
The half-width of the strip of pixels used in the analysis was $17''$  (Blue Filament) and $23''$ (Red Filament).
The origin of offsets along the filaments was taken at the position of the intersection of the two slices, 
\RA{00}{36}{49}{30},
\DECf{+63}{28}{45}{0}.
Position offsets increase westwards (Blue Filament) and north-westwards, (Red Filament), from $-40''$ to $+30''$. 
The position of the cluster of embedded YSOs corresponds to a position offset of approximately $+20''$ for both slices.

\subsubsection{Velocity dispersion}

\begin{figure}[htb]
\resizebox{\hsize}{!}{\includegraphics{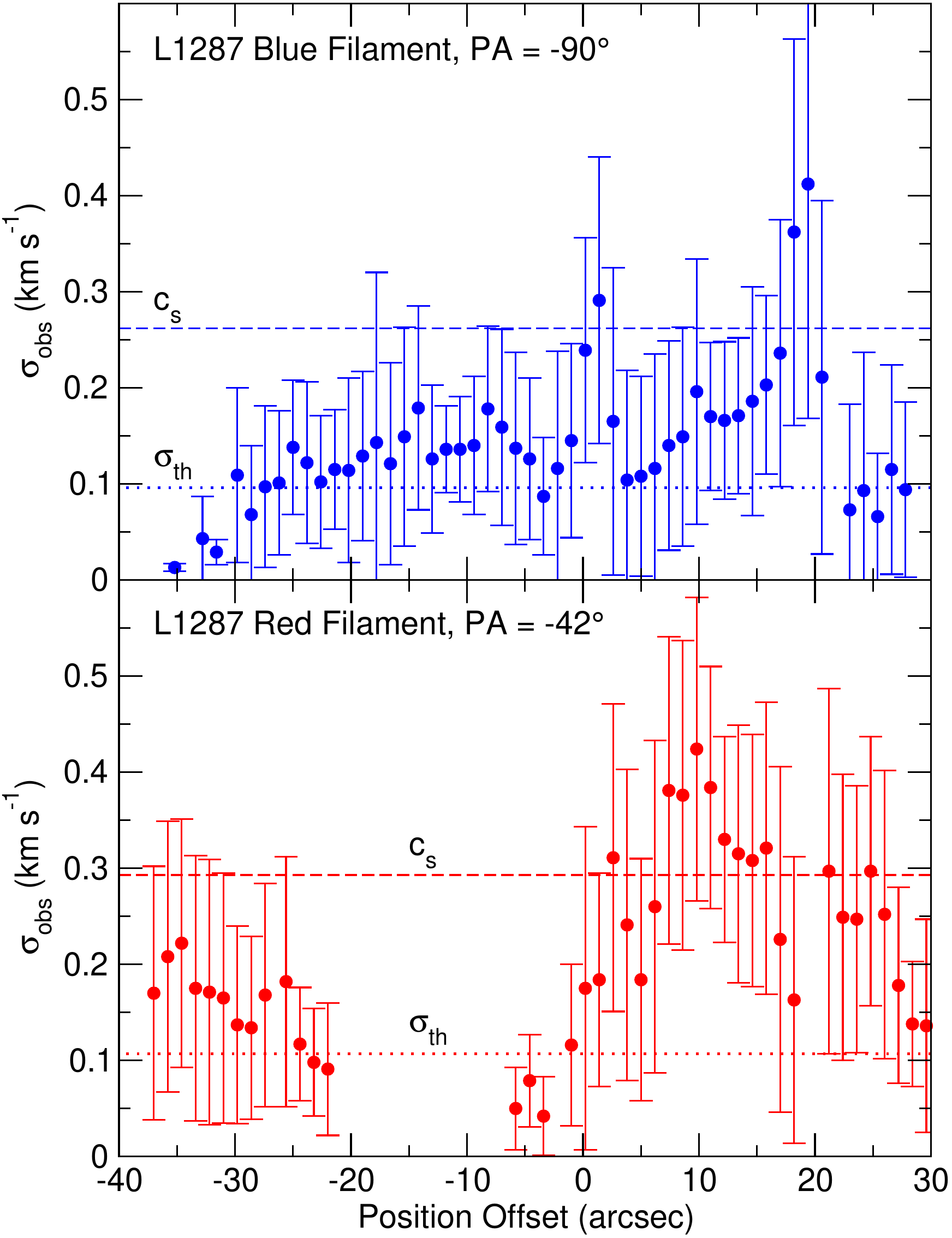}}
\caption{
\nh{} velocity dispersion, $\sigma_V=\Delta V/(8\ln2)^{1/2}$, as a function of position along the Blue Filament (blue dots) and the Red Filament (red dots). 
Each data point is the median value of the velocity dispersion for the points across the filament for a given position along the filament, and the error bars indicate the median of the uncertainty of the values of the velocity dispersion for the same points.
The dashed lines indicate the average sound speed, $c_s$, of the filaments,
0.26 \kms\ (Blue Filament) and 
0.29 \kms\ (Red Filament), 
and the dotted lines indicate average thermal width (see text),
0.10 \kms\ (Blue Filament) and 
0.11 \kms\ (Red Filament), 
The slices used are the same as in Fig.\ \ref{fslices}.
\label{fslice_dv}}
\end{figure}

In Fig.\ \ref{fslice_dv} we show a plot of the velocity dispersion for the slices
of the Blue and Red Filament. 
The increase in velocity dispersion is noticeable for positive offsets in both filaments, which correspond to the position near the cluster of embedded YSOs.

At position offset around zero there is also a sharp increase in line width of the Blue Filament. 
These positions correspond to the superposition of both filaments (the two slices cross at offset 0). The poor spectral resolution of the observations makes difficult to disentangle the emission of both filaments
when they are superposed. This could be a factor that could produce an spurious
increase of the line width, due to the confusion of the emission of both
filaments; or it could be a real effect caused by the mechanical interaction of
the two filaments (see next section). Observations with a higher spectral resolution could help to
see whether this increase in line width is real or not.

\subsubsection{Gas temperature}

\begin{figure}[htb]
\resizebox{\hsize}{!}{\includegraphics{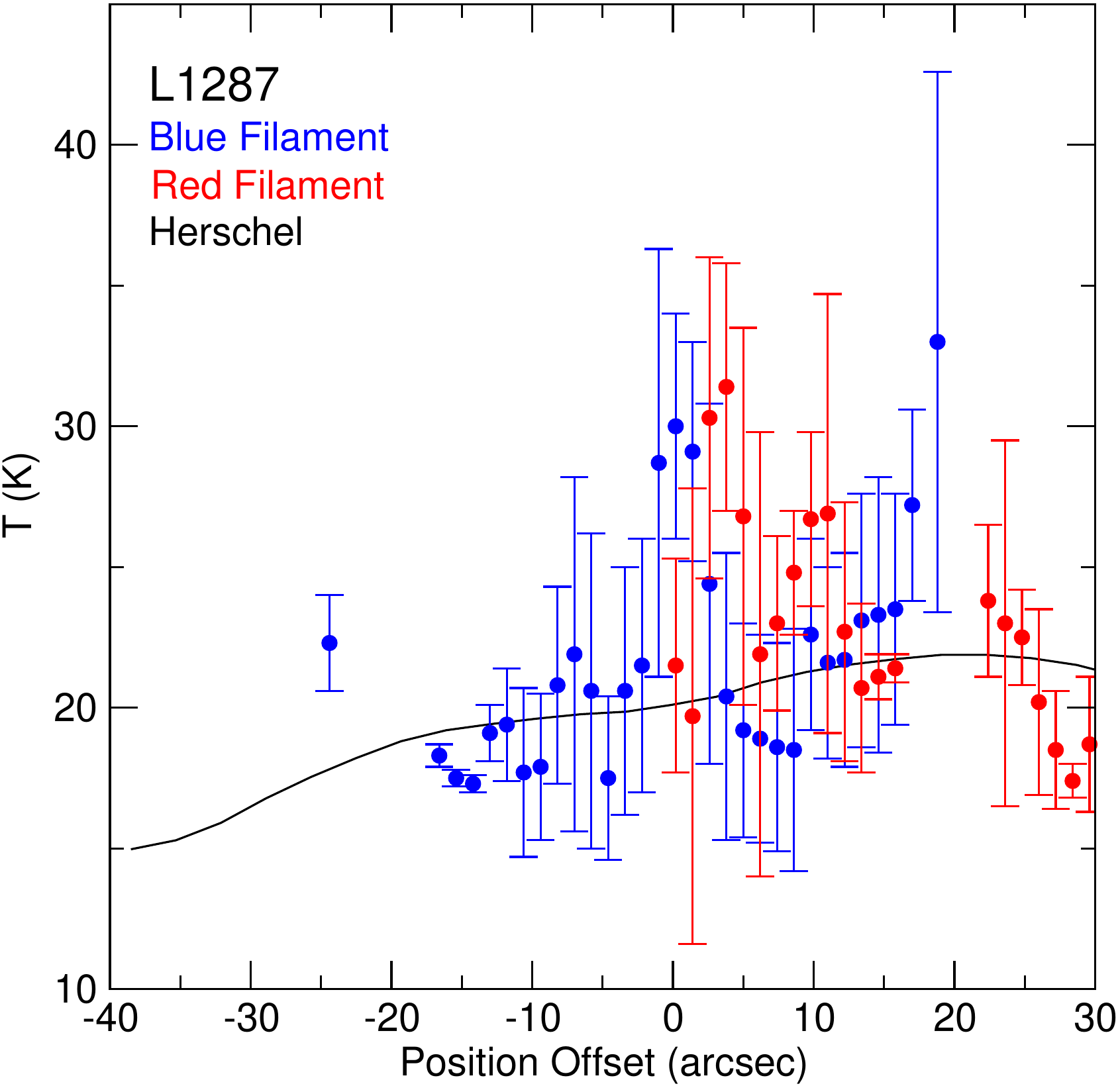}}
\caption{
Kinetic temperature derived from the \nh{} data as a function of position along the Blue Filament (blue dots), the Red Filament (red dots).
Each data point is the median value of the kinetic temperature for points across the filament for a given position along the filament.
The error bars indicate the the median of the uncertainty.
The dust temperature from \Herschel{} data \citep{Jua19}  (beam size $37''$), for a slice encompassing both filaments is shown as a black continuum line. 
\label{fslice_tk}}
\end{figure}

\begin{figure}[htb]
\resizebox{\hsize}{!}{\includegraphics{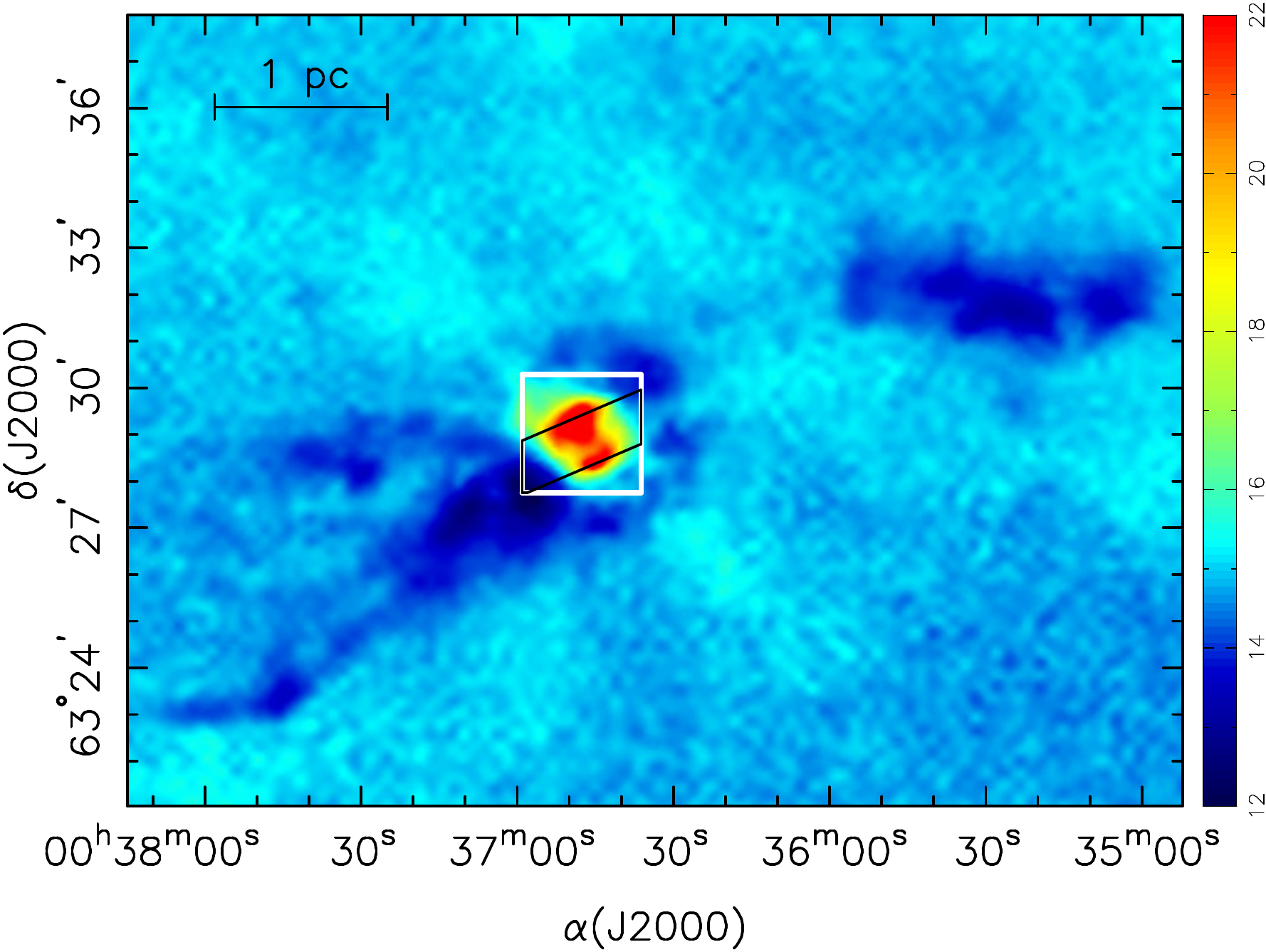}}
\caption{
\Herschel{} large scale map of the dust temperature \citep{Jua19}. 
Note the sharp temperature enhancement in the region mapped in \nh{} (white box).
The slanted black box indicates the slice used to compute the dust temperature from the \Herschel{} data.
\label{fherschelTd}}
\end{figure}

The value of the kinetic temperature for the two slices of the Blue and Red Filament is shown in Fig.\ \ref{fslice_tk}. 
The kinetic temperature could only be derived for pixels of the slices with both \nh{} $(1,1)$ and $(2,2)$ measurements.
The median value of the kinetic temperature is low, 
$T_\mathrm{k} = 22.1$ K for the Blue Filament, and 
$T_\mathrm{k} = 22.9$ K for the Red Filament,
but significantly higher than the median temperature of the large-scale filament derived in the Hi-GAL catalogue, $\sim13.56$ K \citep{Sch20}.
In Fig.\ \ref{fherschelTd} we show the dust temperature map from \Herschel{} \citep{Jua19}. In order to compare the gas and dust temperatures, we computed from the \Herschel{} data the dust temperature for a slice encompassing both filaments (see Fig.{} \ref{fherschelTd}). 
As can be seen in Fig.\ \ref{fslice_tk}, the dust temperature follows closely the variations in gas temperature, smoothed to the \Herschel{} angular resolution ($37''$). 
This agreement shows that the gas and dust are well coupled in L1287.

Two features are apparent, 
(i) the increase in temperature around offset zero, at the crossing of the two filaments, indicating that probably the material of the filaments is interacting with each other (i.e. the superposition of the filaments is real, and not only seen in projection); and 
(ii) the significant increase in temperature at positive offsets at $\sim20''$, at the position of the cluster of embedded YSOs, suggesting gas heating caused by the interaction with the the high-velocity gas of the outflows driven by IRAS 00338+6312/VLA 3 and by RNO 1C/VLA 1 (see Fig.\ \ref{hfs_trot_zoom}).
An alternative gas heating mechanism is radiative heating from the radiation of the embedded cluster. 
A simple application of the Stephan-Boltzmann law gives that an isotropic luminosity of $\sim20$ \Lsol{} at a distance of 6\,000 au (the radius of the cavity), can produce a heating at the walls of the cavity up to a temperature of $\sim20$ K.
Thus, both mechanical and radiative heating could contribute to the observed gas temperature increase.

\subsubsection{Mach number}

\begin{figure}[htb]
\resizebox{\hsize}{!}{\includegraphics{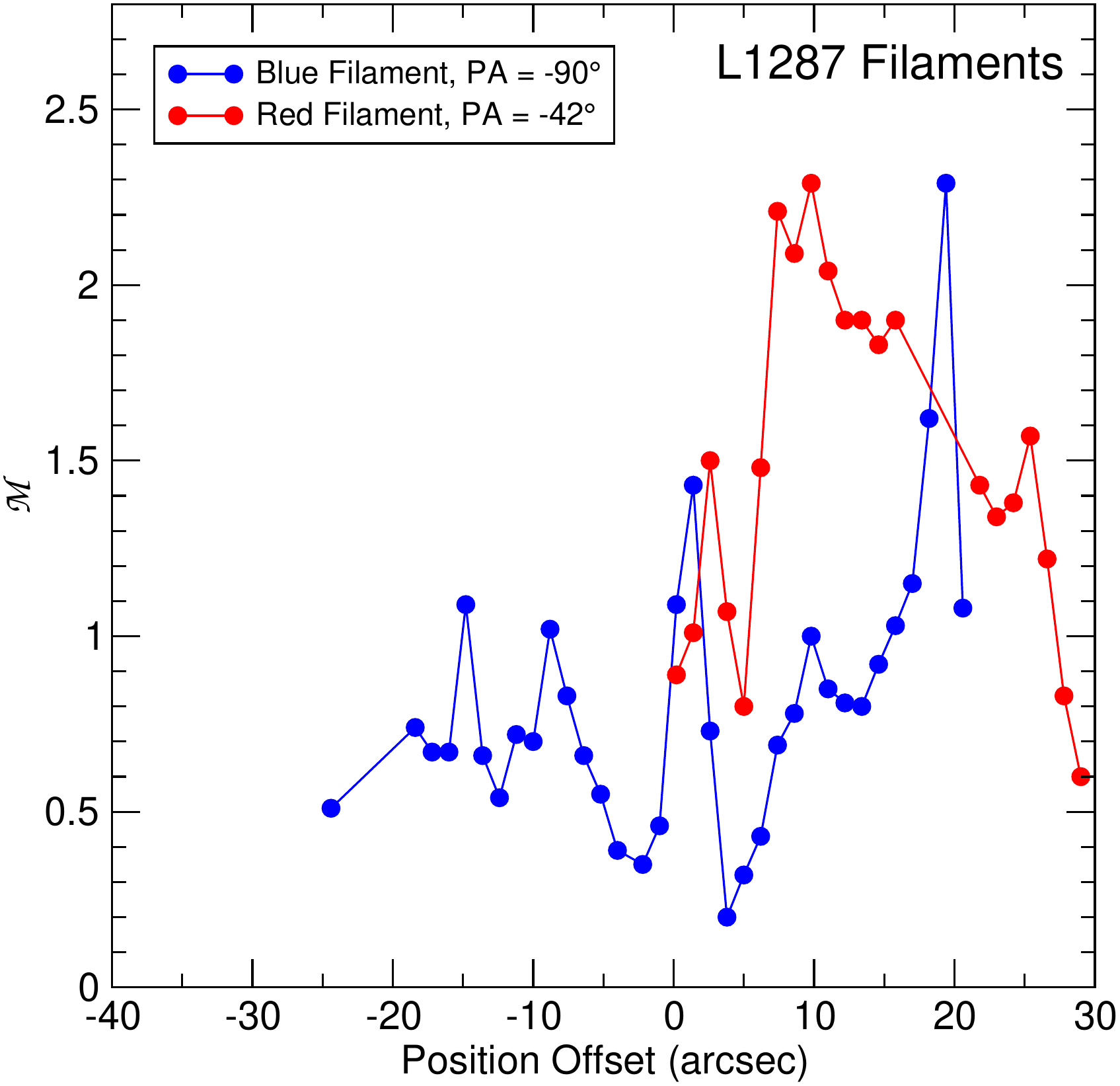}}
\caption{
Median value of the Mach number derived from the \nh{} data, 
as a function of position along the Blue Filament (blue dots) and 
the Red Filament (red dots). 
The slices used are the same as in Fig.\ \ref{fslices}.
\label{fslice_mach}}
\end{figure}

In order to characterize the importance of turbulence in the filaments, we estimated the median value of the Mach number, ${\cal M}$, across the filament as a function of position along the filament. The Mach number is calculated as in
\citet{Pal15}
\begin{equation}
{\cal M}=\frac{\sigma_\mathrm{3D,nth}}{c_s}, 
\end{equation}
where $c_s$ is the isothermal sound speed,
\begin{equation}
c_s= \left(\frac{k T_\mathrm{k}}{\mu_\mathrm{gas}m_\mathrm{H}}\right)^{1/2},
\end{equation}
being
$k$ the Boltzmann constant,
$T_\mathrm{k}$ the gas kinetic temperature,
$\mu_\mathrm{gas}=2.3$, the average molecular mass of the interstellar gas,
$m_\mathrm{H}$ the mass of the hydrogen atom,
and
$\sigma_\mathrm{3D,nth}$ is the 3-D non-thermal velocity dispersion,
\begin{equation}
\sigma_\mathrm{3D,nth}=\sqrt{3}\,\sigma_\mathrm{1D,nth}.
\end{equation}
The 1-D non-thermal velocity dispersion, $\sigma_\mathrm{1D,nth}$, is given by the (squared) difference between observed and thermal velocity dispersion,
\begin{equation}
\sigma_\mathrm{1D,nth}=
\left(\sigma_\mathrm{1D,obs}^2-\sigma_\mathrm{th}^2\right)^{1/2},
\end{equation}
with $\sigma_\mathrm{1D,obs}$ given by the observed line full-width at half-maximum (deconvolved from the channel width), $\Delta V$,
\begin{equation}
\sigma_\mathrm{1D,obs}= \frac{\Delta V}{(8\ln2)^{1/2}},
\end{equation}
and $\sigma_\mathrm{th}$ being the thermal line width,
\begin{equation}
\sigma_\mathrm{th}=
\left(\frac{kT_\mathrm{k}}{\mu_\mathrm{NH_3}m_\mathrm{H}}\right)^{1/2},
\end{equation}
where
$\mu_\mathrm{NH_3}=17$ is the molecular mass of the \nh{} molecule. 

The results are shown in Fig.\ \ref{fslice_mach} as blue dots (Blue Filament) and red dots (Red Filament).
The motion of both filaments is nearly subsonic,  except for the positions along the slices at positive offsets at $\sim10''$ or $20''$,  at the position of the cluster of embedded YSOs, where the Mach number reaches values of $\sim2$ (moderately supersonic) indicating the effect of the interaction with the high-velocity gas of the outflows driven by the IRAS source and RNO 1C.
Alternatively, the larger Mach numbers near the cluster of YSOs could be indicative of on-going infall from larger scales, consistent with theoretical models \citep[e.g.,][]{Vaz19}.

\subsubsection{Mass per unit length}
\label{sec_slice_massl}

\begin{figure}[htb]
\resizebox{\hsize}{!}{\includegraphics{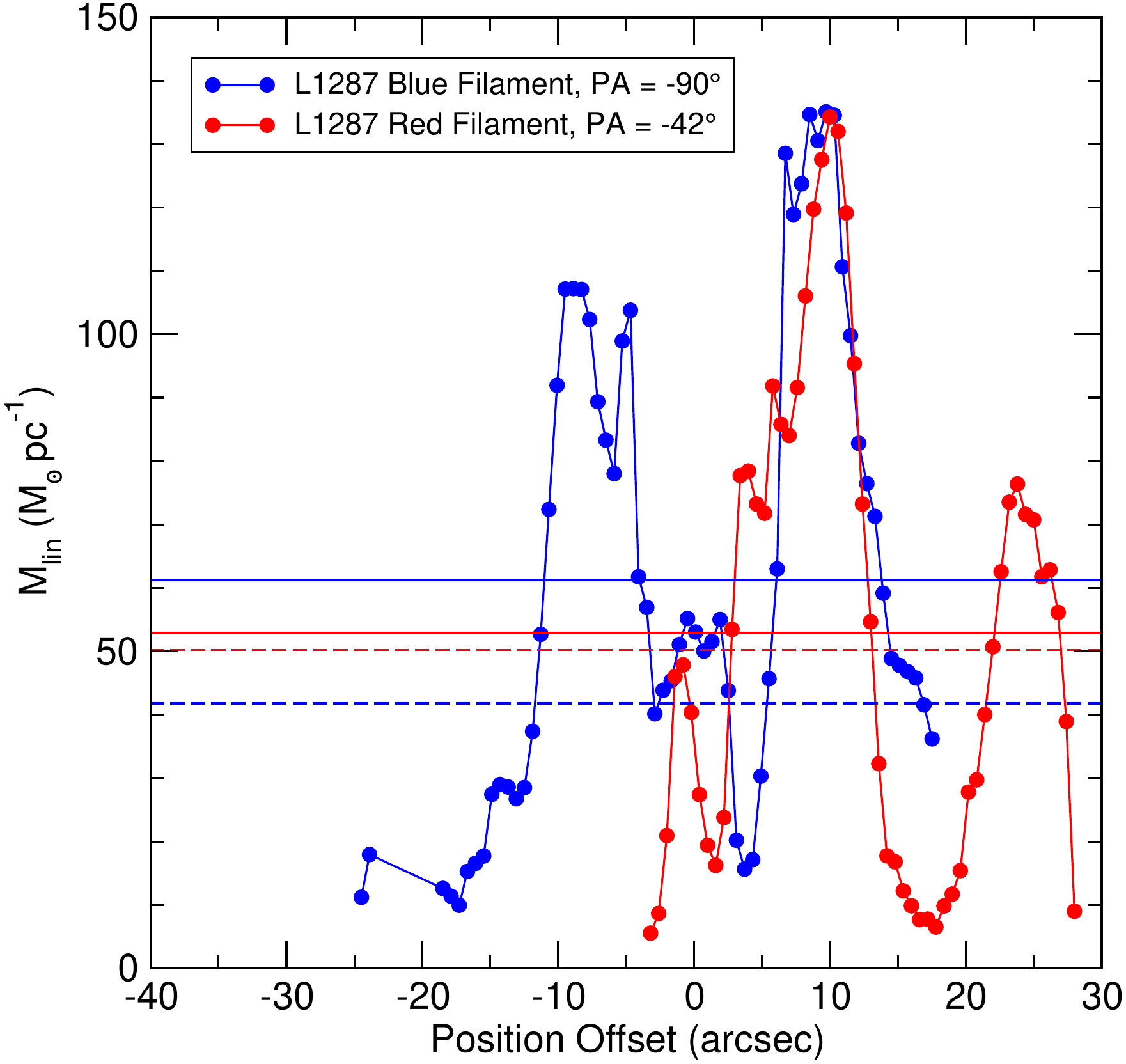}}
\caption{
Mass per unit length as a function of position along the Blue Filament (blue dots) and the Red Filament (red dots).
The mass per unit length has been obtained from the sum of \nh{} column densities across the filament, for each position along the slice.
The horizontal solid lines indicate the average value for the whole filament, 
61.2 $M_\odot$~pc$^{-1}$ (Blue Filament), and 
53.0 $M_\odot$~pc$^{-1}$ (Red Filament).
The dashed lines indicate the critical mass per unit length for thermal and non-thermal support}, $2\sigma_\mathrm{tot}^2/G$ (Table \ref{tfilament}),
42.8 $M_\odot$~pc$^{-1}$ (Blue Filament), and 
50.2 $M_\odot$~pc$^{-1}$ (Red Filament).
The slices used are the same as in Fig.\ \ref{fslices}.
\label{fslice_massl}
\end{figure}

\begin{table}[htb]
\centering
\caption{Blue and Red Filament stability analysis
\label{tfilament}}
\begin{tabular}{lccccc}
\hline\hline
&
$c_s$ &
$\sigma_\mathrm{tot}$\tablefootmark{a} &
$\Ml$\tablefootmark{b} & 
$\Mlc$\tablefootmark{c} & 
 \\ 
Fil. & 
(\kms) & 
(\kms) & 
(\Msolpc) & 
(\Msolpc) & 
$\alpha_v$\tablefootmark{d} \\ 
\hline
Blue   & 0.28 & 0.30 & 61.2 & 41.8 & 0.7 \\
Red    & 0.29 & 0.33 & 53.0 & 50.2 & 0.9 \\
\hline
\end{tabular}
\tablefoot{ 
\tablefoottext{a}
{Total velocity dispersion of the interstellar gas (see text).}
\tablefoottext{b}
{Average value for the whole filament of the mass per unit length $\Ml$.}
\tablefoottext{c}
{Critical value of mass per unit length for thermal and non-thermal support, $\Mlc= 2\sigma_\mathrm{tot}^2/G$.}
\tablefoottext{d}
{Virial factor, $\alpha_v= \Mlc/\Ml$.}
}
\end{table}

An important parameter of an elongated structure as a filament is its mass density distribution, i.e.{} the mass per unit length. 
For computing this we used the column density determined for the pixels of the slices of the Blue and Red Filaments. 
The mass per pixel spacing was obtained as the sum of the \nh{} column density across the filament, for each position along the slice, converted to total mass using an \nh{} abundance $X(\mathrm{NH_3})=1\times10^{-8}$, and divided by the size of the pixel in pc.
The results are shown in Fig.{} \ref{fslice_massl}.
The solid lines indicate the average value for the whole filament, $\Ml=61$ \Msolpc\ (Blue Filament) and $\Ml=53$ \Msolpc\ (Red Filament) (see Table \ref{tfilament}). 
The typical values (50\% of the points) obtained for the Blue Filament are $30\mbox{~\Msolpc}\le \Ml \le91$ \Msolpc.
The typical values for the Red Filament are in general a bit lower, $18\mbox{~\Msolpc}\le \Ml \le76$ \Msolpc.

The stability of a self-gravitating filament can be studied  through the virial parameter, $\alpha_v$,
\begin{equation}
\alpha_v = \frac{\Mlc}{\Ml} 
\end{equation}
where the critical value of the mass per unit length, $\Mlc$, for an infinite length cylinder is given by (see Appendix \ref{amodel})
\begin{equation}\label{eqsigma}
\Mlc=\frac{2\sigma_\mathrm{tot}^2}{G},
\end{equation}
and is the mass per unit length whose self-gravitation is balanced by the thermal and non-thermal motions of the gas, given by the total velocity dispersion $\sigma_\mathrm{tot}$. 
The total velocity dispersion is evaluated as in \citet{Pal15}
\begin{equation}
\sigma_\mathrm{tot} = \left(c_s^2 + \sigma_\mathrm{1D,nth}^2\right)^{1/2}.
\end{equation}
For $\alpha_v<1$ the cylinder cannot support its own mass against gravitation.
It is collapsing, or there are other mechanisms of support (e.g., magnetic field) not taken into account. 
For $\alpha_v>1$ the cylinder is expanding and will dissipate, unless there are other mechanisms (external pressure) to confine its material.

The average values of $\sigma_\mathrm{tot}$, $\Mlc$, and $\alpha_v$ for the pixels of the slices of the Blue and Red Filaments are given in Table \ref{tfilament}. As can be seen from this analysis, both filaments are near virial equilibrium ($\alpha_v=0.7$ for the Blue Filament, and $\alpha_v=0.9$ for the Red Filament). 
Or, in other words, the mass that can be supported with the observed total velocity dispersion is similar to the mass per unit length observed for both filaments. However, this support must come from both thermal and non-thermal motions, because the critical mass obtained from the thermal motions only (that is, using $c_s$ instead of $\sigma_\mathrm{tot}$), is only $\sim40$~\Msolpc, below the observed values (Table \ref{tfilament}).

\subsubsection{Self-gravitating isothermal cylindrical filament model
\label{sect_filamentmodel}
} 

The self-gravitating isothermal cylindrical filament model (see Appendix \ref{amodel}) predicts a column density as a function of distance from the filament axis $p$ that is a power law with index $-3$, flattened at the origin (see Equation \ref{eqnp}).
The model depends on two parameters only,  the column density at the filament axis, $N_0$,  and the characteristic radius, $h$, distance for which the column density is 0.35 times $N_0$.

\begin{figure}[htb]
\resizebox{\hsize}{!}{\includegraphics{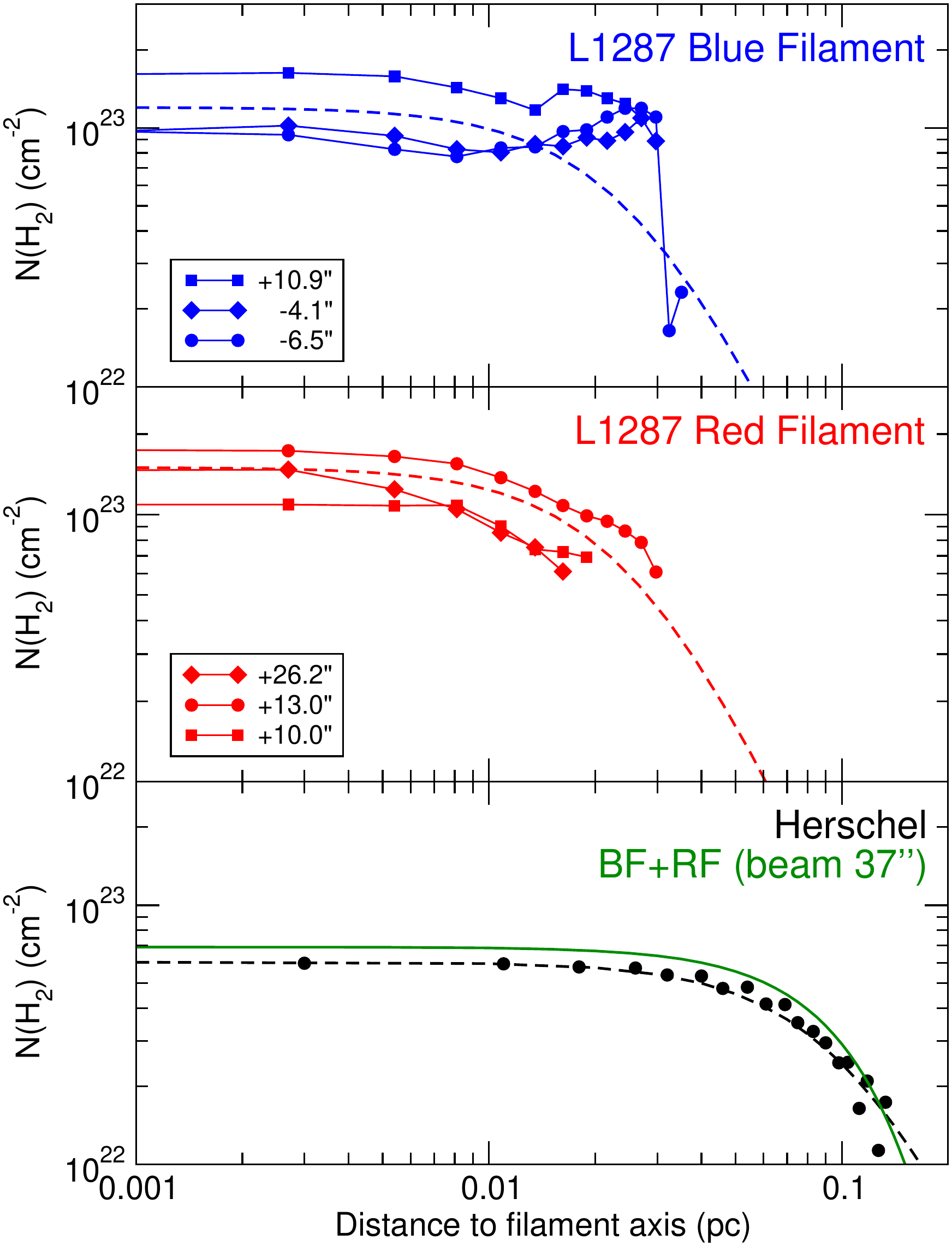}}
\caption{
Hydrogen column density as a function of distance to the filament axis, for three positions along the Blue Filament, at position offsets $-6\farcs5$, $-4\farcs1$, $+10\farcs9$ (top panel), and along the Red Filament, at position offsets $+10\farcs0$,  $+13\farcs0$, $+20\farcs2$ (middle panel). 
The axes of the filaments have been taken as the central lines of the slices used in Fig.\ \ref{fslice_vlsr}.
The blue and red dashed lines represent the self-gravitating isothermal filament models fitted for the Blue and Red Filament data (see Table \ref{tisothermal}).
The bottom panel shows the \hh\ column density from \Herschel{} \citep{Jua19} (black filled circles) and the self-gravitating isothermal filament model fit to the \Herschel{} data (black dashed line).
The green continuum line is the sum of the models for the Blue and Red
Filaments, convolved with the \Herschel{} beam ($\mathrm{FWHM}=37''$). 
\label{fslice_nh}}
\end{figure}

We obtained the \hh\ column density as a function of distance to the filament axis for three positions of the Blue Filament and three other positions for the Red Filament, for which the column densities were estimated up to a distance $\sim0.03$ pc. 
For larger distances the \nh $(1,1)$ or $(2,2)$ emission was too faint to fit the spectra, and the column density could not be calculated. 
The results obtained are shown in Fig.\ \ref{fslice_nh}.
As can be seen, for the Red Filament the column density decreases with increasing distance. For the Blue Filament this behavior is not so clear, but we can see that in one position we were able to see a significant drop in column density with distance. 
The explanation for this could be that the Blue Filament is not a single entity, but there are hints of a substructure of subfilaments, or threads, that are not discernible with the present sensitivity, angular resolution, and spectral resolution of the \nh{} observations.
These subfilaments would be comparable to the velocity-coherent filamentary components (grouped into bundles) reported by \citet{Hac13, Hac17}, \citet{Shi19b}, or \citet{Che20}, which are typically separated about 0.2--0.9 \kms.

\begin{table}[htb]
\centering
\caption{Isothermal filament model for the Blue and Red Filaments
\label{tisothermal}}
\begin{tabular}{l@{\extracolsep{0.9em}}c@{\extracolsep{0.9em}}c@{\extracolsep{0.9em}}c@{\extracolsep{0.9em}}c@{\extracolsep{1em}}c}
\hline\hline
&
$h$\tablefootmark{a} & 
$N_0(\mathrm{H_2})$\tablefootmark{a} & 
$n_0(\mathrm{H_2})$\tablefootmark{b}  &
$\Ml$\tablefootmark{c} &
$\sigma_\mathrm{tot}$\tablefootmark{d} \\
Fil. & 
(pc) &
(cm$^{-2}$) &
(cm$^{-3}$) &
($M_\odot$~pc$^{-1})$ &
(km s$^{-1}$) \\
\hline
Blue & 0.027& $1.2\times10^{23}$ & $9.1\times10^5$ & 146 & 0.56 \\
Red  & 0.025& $1.5\times10^{23}$ & $1.3\times10^6$ & 167 & 0.60 \\
\hline
\end{tabular}
\tablefoot{ 
\tablefoottext{a}{Characteristic radius, $h$, and column density at the filament axis, $N_0(\mathrm{H_2})$, (see text) obtained from the fit to the observed column density as a function of distance to the filament axis.} 
\tablefoottext{b}{Volume density at the filament axis derived from the model, $n_0= 2N_0/\pi h$.}
\tablefoottext{c}{Mass per unit length derived from the model, $\Ml=2\mu_\mathrm{H_2}m_\mathrm{H}N_0h$.}
\tablefoottext{d}{Total velocity dispersion derived from the model, $\sigma_\mathrm{tot}=(G\mu_\mathrm{H_2}m_\mathrm{H}N_0h)^{1/2}$.}
}
\end{table}

We performed a least-squares fit of the isothermal cylindrical model to the column density as a function of the distance to the filament axis, up to a distance of $\sim0.03$ pc, for both filaments. 
The parameters of the best-fit models are given in Table \ref{tisothermal}, and the models are also shown in Fig.{} \ref{fslice_nh}. 
As can be seen, the fit is very good for the Red Filament, and fairly good for the Blue Filament. 
The models fitted for the Blue and Red Filaments are very similar.
The characteristic radii obtained are $\sim0.03$ pc for both filaments. 
This value is different from the typical value of 0.1 pc found for \Herschel{} filaments by \citet{Arz11}, but consistent with values found in the Integral Shaped Filament of Orion \citep{Hac18}, and in Serpens and Perseus \citep{Dha18}.
The volume density deduced from the model (Eq.{} \ref{eqnp}) is $\sim1\times10^6$ \cmt, and the mass per unit length predicted by the models is $\sim150$ \Msolpc.
The total velocity dispersion of the interstellar gas needed to have a stable isothermal filament (Eq.\ \ref{eqsigma}) is $\sim0.58$ \kms.
In general, we can say that the isothermal cylindrical model fits the observed column density as a function of distance to the filament axis.

We compared the column density obtained from the ammonia data and from the dust continuum emission from \Herschel{}. 
The comparison can be seen in the bottom panel of Fig.\ \ref{fslice_nh}, where we show the column density across the filament from the \Herschel{} data \citep{Jua19}, and the sum of column densities of the isothermal cylindrical models for the Blue and Red Filaments, convolved with the \Herschel{} beam ($\mathrm{FWHM}=37''$). 
There is a clear agreement between the \nh{} and \Herschel{} data. This is indicative that the ammonia abundance used, $X(\mathrm{NH_3})=1\times10^{-8}$, is in agreement with the \Herschel{} assumptions made in the data reduction \citep{Jua19},
and that we are detecting the bulk of the filaments mass for distances up to $p\simeq0.03$ pc from the filament axis.

However, there are discrepancies between the values obtained from the observations (see Table \ref{tfilament}) and from the model (see Table \ref{tisothermal}) of the mass per unit length and the velocity dispersion.
The apparent discrepancy in velocity dispersion can be explained by the fact that the \nh{} observations are tracing small scales of the order of 0.03 pc (that is, of the order of the characteristic radius $h$), while the model takes into account the behavior of the gas at a large scale (in fact, up to an infinite distance from the filament axis). The velocity dispersion measured at large scales will be larger than the value measured at small scales from the \nh{} emission. For instance, for the Blue Filament (see next Section), we measure a velocity gradient across the filament of 11 \kmspc{}. Such a velocity gradient, measured at a scale of 0.15 pc 
(that is, a diameter corresponding to a radius of only 2.5 times the characteristic radius $h$), 
would produce an increase in the velocity dispersion of 0.5 \kms{}, making a total velocity dispersion of 0.6 \kms{}, in agreement with the self-gravitating model. 

Regarding the discrepancy in the mass per unit length, the column density obtained from  the \nh{} observations could only be integrated up to a distance of the order of the characteristic radius of the filaments, and the model predicts that the mass per unit length up to the characteristic radius $h$ is half the total mass of the filament (see Appendix \ref{amodel}).
Thus, the discrepancy in the masses is smaller, although  there is still a deficit of a 16\% in the measured mass of the Blue Filament, and of a 37\% in that of the Red Filament. 
This may be indicative that the \nh{} observations were not able to recover all the mass of the filaments because we were not sensitive to the cold and low-density component of the gas of the filament, for which the \nh{} $(1,1)$ or $(2,2)$ are not excited.

\subsubsection{Central velocity: velocity gradients along and across the filaments}
\label{sec_slice_vlsr} 

\begin{figure}[htb]
\resizebox{\hsize}{!}{\includegraphics{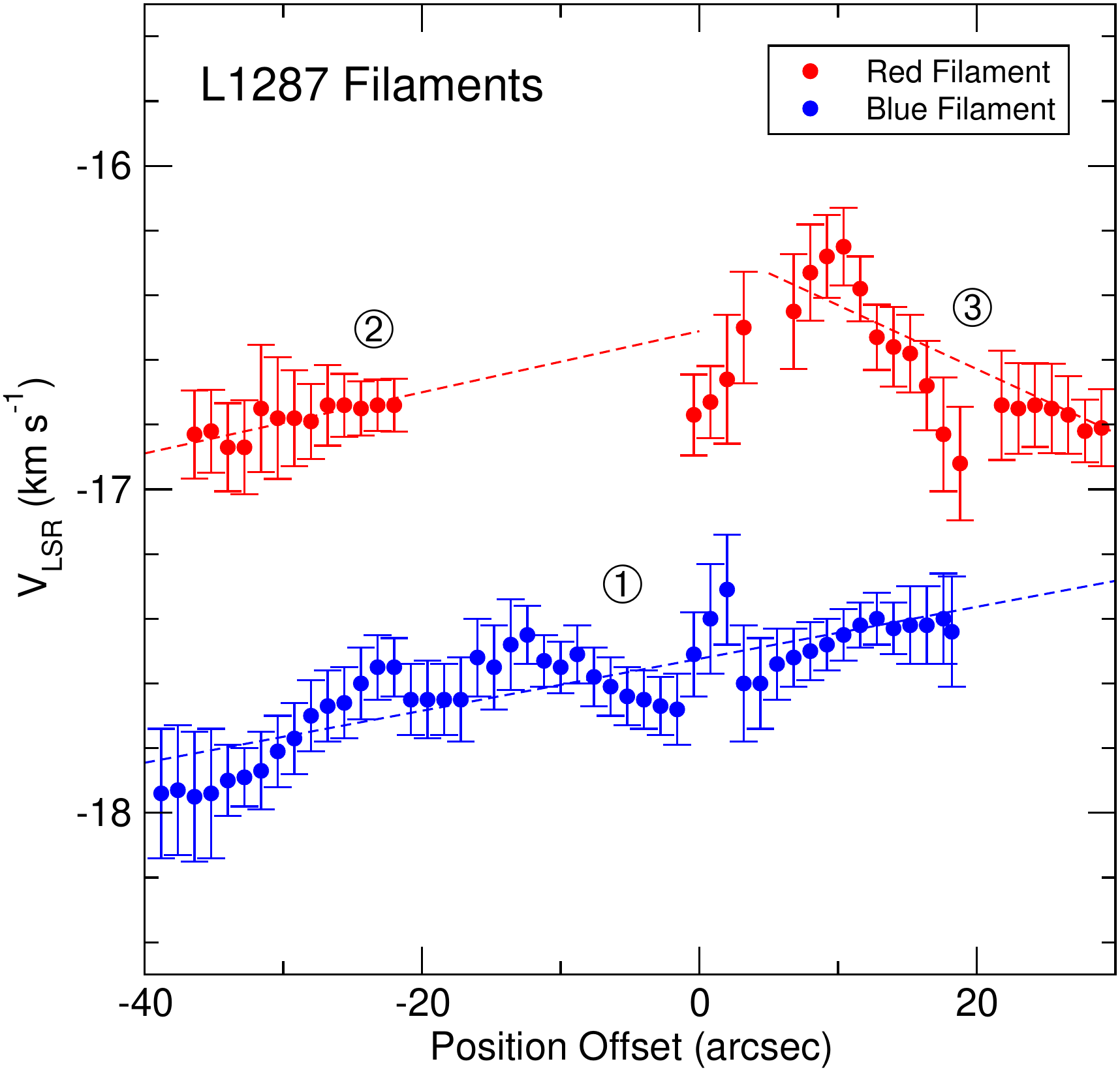}}
\caption{
Median of the central velocity of the \nh{} $(1,1)$ line for points across the filament, as a function of position along the Blue Filament (blue dots) and the Red Filament (red dots). 
The two slices used are the same as in Fig.\ \ref{fslices}.
The error bars indicate the median of the uncertainty of the values of the central velocity. 
The range of values is typically three times the median uncertainty.
The dashed lines indicate the velocity gradients along the filaments. They are identified with labels for the Blue Filament (``1'') and the Red Filament (``2'' and ``3''). The same labels are used in the sketch of Fig.\ \ref{fsketch}.
\label{fslice_vlsr}}
\end{figure}

The $(1,1)$ central velocities for the slices of the Blue and Red Filament are shown in Fig.\ \ref{fslice_vlsr}. 
The Blue Filament, for position offsets between $-40''$ and $+20''$, shows an increasing velocity gradient of $\sim1.8$ \kmspc{} (labeled as ``1'' in Fig.\ \ref{fslice_vlsr}).
For the Red Filament there is an increasing velocity gradient of $\sim2.1$ \kmspc{} for offsets from $-40''$ to $+10''$ (labeled as ``2''), 
and a decreasing velocity gradient of $\sim4.4$ \kmspc{} between offsets $+10''$ and $+30''$ (labeled as ``3''), encompassing the location of the cluster of YSOs, at an offset of $+20''$.
There is a hint of an oscillatory pattern in the velocity gradient seen along the Blue Filament, mainly around offsets $-20''$ and $0''$. These offsets are coincident with the position of cores embedded in the Blue Filament (see Fig.{} \ref{fslices}). 
Thus, these oscillations could originate from gas motions flowing towards the cores being formed within the filament, consistent with the model proposed by \citet{Hac11}. 
Higher spectral resolution observations would definitely help establish the significance of such oscillations.

\begin{figure}[htb]
\resizebox{\hsize}{!}{\includegraphics{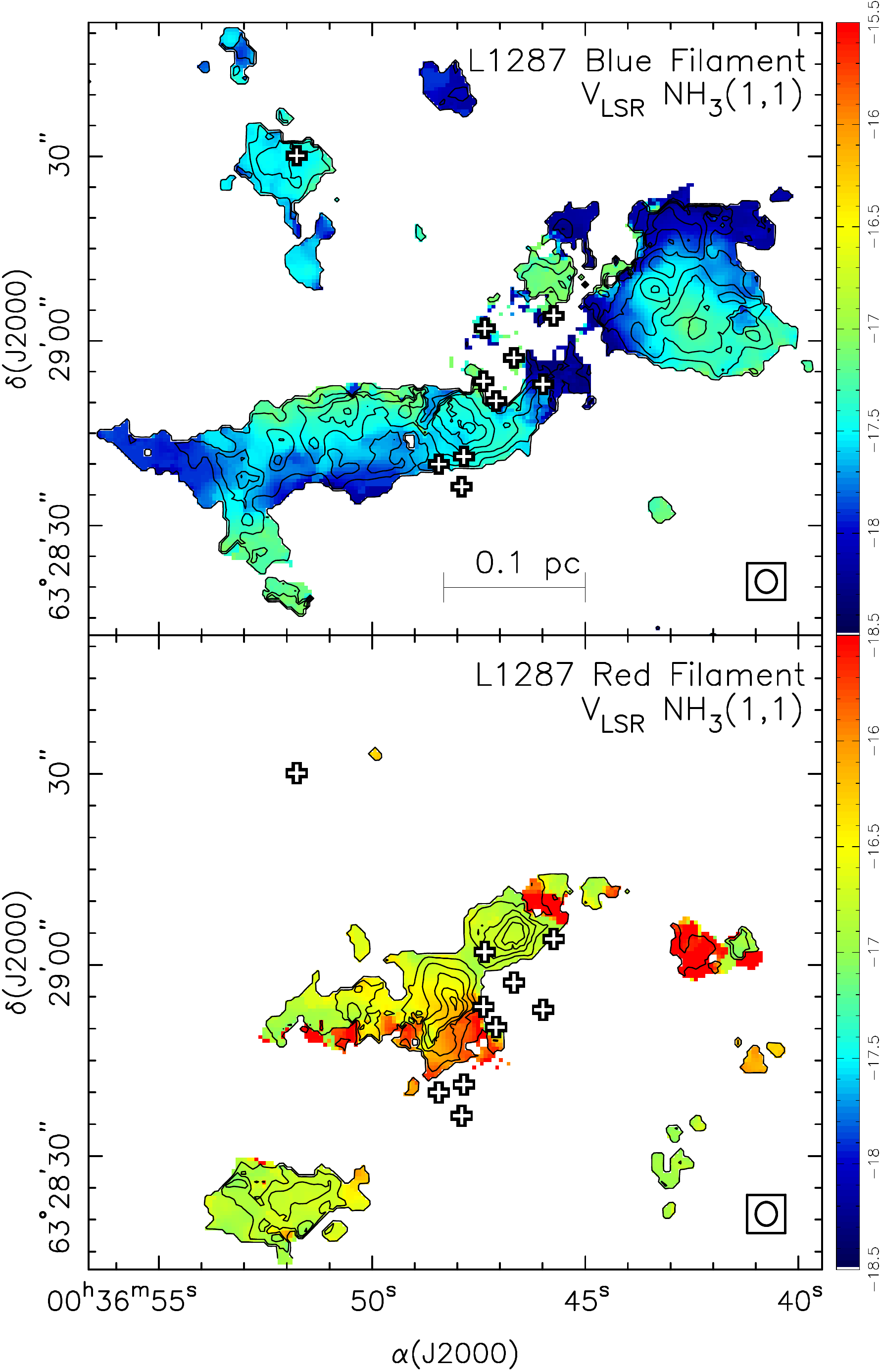}}
\caption{
Central velocity of the \nh{} $(1,1)$ line of the L1287 Blue and Red Filaments.
The color scale for both panels ranges from $-18.5$ \kms{} to $-17.1$ \kms{}.
The contours are the $(1,1)$ integrated line intensity of Fig.\ \ref{hfs_ataumdv}.
The crosses correspond to YSOs identified in Fig.\ \ref{caha}.
\label{f_vlsrcomp2}}
\end{figure}

Figure~\ref{f_vlsrcomp2} displays the central velocity map of the Blue and Red Filaments. 
While the Red Filament does not show any clear velocity pattern, the Blue Filament shows a velocity gradient across its entire width, with a value of $\sim11$ \kmspc{}, significantly larger than the velocity gradient observed along the filament. 
Evidence of such strong velocity gradients across the filaments has been found in other star-forming regions, e.g.,
Serpens South \citep{Fer14,Dha18},
IRDC 18223 \citep{Beu15},
SDC 13 \citep{Wil18}, and 
NGC 1333 \citep{Dha18}.
Interestingly, the clump located to the west of the main part of the Blue Filament also shows a velocity gradient across the filament axis of $\sim8$~\kmspc{}, but in the opposite direction. 

The observed velocity gradient  of $\sim11$ \kmspc{} across the Blue Filament could result from a series of velocity-coherent subfilaments with slightly different velocities that overlap on the plane of the sky and are not resolved by our data. 
This is consistent with the flattened column density profile found for the Blue Filament (Fig.{} \ref{fslice_nh}) and has been found in other filaments in the literature \citep{Hac13, Hac17, Shi19b}.

Alternatively, the velocity gradient could be due to infalling material onto the filament, as has been claimed in other cases \citep{Fer14, Gon18, Shi19a, Che20}.
Although in Section \ref{sect_filamentmodel} we presented a model of a self-gravitating hydrostatic isothermal cylindrical filament, this is a very idealized case which is not taking into account ingredients such as accretion onto the filament and further contraction. We consider here a contracting filament.
In this scenario, the free-fall velocity and radius allow us to derive the free-fall time (Eq.\ \ref{eqc17}). 
For a velocity gradient of 11 \kmspc{}, a radius of the filament of 0.027~pc, and a mass per unit length of 61.2 \Msolpc{} (Table \ref{tfilament}), the free-fall velocity is $\sim0.3$ \kms{} and the free-fall time is $\sim0.05$ Myr.
This time has to be compared with the age of the filament.
This can be estimated to be $\sim2$~Myr, based on the evolutionary stage of the embedded YSOs, which range from a Class~0 object (IRAS\,00338+6312/VLA\,3) to a Class~I/II object (FU Ori system RNO\,1B/1C), and hence the age estimate from prestellar to Class~0 and Class~II objects is $\sim0.6$~Myr and 2~Myr, respectively \citep{Eva09}.
Thus, we can conclude that the accretion would take place in a time scale much shorter that the age of the filament and hence, the accretion scenario is compatible with the observed velocity gradient across the filament.
We note that in Appendix \ref{atff}, a cylindrical filament has been considered. However, in order to observe the velocity gradient across the filament, a slightly flattened filament would be required.

A second possible scenario that could produce the velocity gradient across the filaments is rotation \citep[see, for instance, ][]{Gon19}. 
We used Eq.\ (1) of \citet{Rec14} to roughly estimate the contribution of rotational energy compared to gravity. 
It is worth emphasizing the approximate nature of the adopted energy balance, as it does not include the effects of a radial density profile and the presence of internal temperature gradients.
The observed velocity gradient corresponds to an angular velocity of $\omega\simeq3.6\times10^{-13}$~s$^{-1}$.  Adopting the mass per unit length and the characteristic radius given in Tables \ref{tfilament} and \ref{tisothermal} ($\Ml=61.2$ \Msolpc{}, $R_{c}\simeq0.027$ pc), we obtained a ratio of rotation to gravitational energy, $T/W\simeq0.17$. 
This suggests that a rotation of the filament could explain the observed velocity gradient, with gravity being the dominant energy, but with a small contribution of the rotational energy that should be taken into consideration in the energy balance of the filament. 

\subsection{A global scenario: Infall, inflow, and outflow in L1287}

In Section~\ref{sec_slice_vlsr} we analyzed the kinematics along the Red and Blue Filaments. In particular, Fig.\ \ref{fslice_vlsr} revealed at least three distinct velocity gradients, two associated with the Red Filament and one associated with the Blue Filament. 
One possible interpretation for a velocity gradient along a filamentary structure is that it is tracing inflow motions towards a deep potential well. In the case of L1287 the potential well could be produced by the cluster of YSOs.
This is consistent with simulations of formation of filaments in molecular clouds that undergo collapse \citep[e.g., ][]{Gom14}, and with previous observations of kinematics in filamentary structures in
Serpens \citep{Kir13, Lee14, Gon18, Dha18},
TMC 1 \citep{Dob19},
several high-mass star-forming cores \citep{Tac14, Lu18},
specific infrared dark clouds (SDC1: \citet[]{Per14}; G333: \citet{Vee18}; G22: \citet{Yua18}; G14: \citet{Che19}),
Mon R2 \citep{Tre19},
Sgr B2 \citep{Sch19}.

\begin{figure}[htb]
\resizebox{\hsize}{!}{\includegraphics{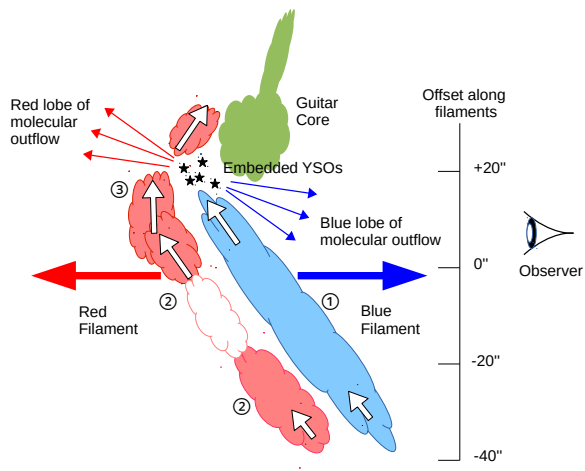}}
\caption{
Sketch of the velocity gradients along the Blue and Red Filaments in L1287. The numbers 1, 2, and 3 identify the velocity gradients along the filaments shown in Fig.\ \ref{fslice_vlsr}.
The scale in arcsec corresponds to the position offset along the filaments of the same figure.
}
\label{fsketch}
\end{figure}

Assuming that this interpretation is correct, it is possible to build a 3D picture of the filamentary structures with respect to the observer, presented schematically in Fig.\ \ref{fsketch}, where each filament portion associated with a velocity gradient is labeled with a number corresponding to the labels of Fig.\ \ref{fslice_vlsr}. 
The increasing velocities observed along the filaments correspond to the velocity gradients ``1'' and ``2'' of Fig.{} \ref{fslice_vlsr}.
The bend in the Red Filament explains the change in the velocity gradient at position offset $+10''$, from ``2'' to ``3''. 
The embedded YSOs are located at position offset $+20''$. 
The gap in the Red Filament is indicated as the part of the Red Filament sketched in white with a red contour. 
The Guitar Core appears offset from the embedded YSOs because it is seen in projection at the position of the YSOs, but with no interaction with them, as discussed above.
In this scenario, the velocity gradients ``2'' and ``3'' correspond to the `redshifted' converging flow detected by \citet{Jua19} in DNC and other molecular tracers. 
Regarding the `blueshifted' converging flow reported in \citet{Jua19}, this cannot be easily identified in our \nh{} data, and it could rather result from material affected by the passage of the blueshifted lobe of the outflow driven by the IRAS source or, alternatively, it could be material associated with the Guitar Core, which is associated with the bluest \nh{} gas found in the region. 
In the sketch, we have also included the Guitar Core, where infall has been detected in Section~\ref{sec_guitar}. 
Therefore, the L1287 region at the center of the 10~pc filamentary structure seems to be associated with infall (Guitar Core), inflow (Blue and Red filaments) and outflow (from the IRAS source especially) simultaneously. 
It seems that the different velocity filamentary structures identified from the \nh{} emission converge at the location of the YSOs and thus our data strongly suggest that the cluster in L1287 is being formed at the intersection of filamentary structures where inflow/infall of material is being funnelled. 
This is fully consistent with the model of formation of clusters presented by \citet{Vaz17} and with a number of observational works \citep[see, for instance,][]{Jim14}, giving further support to a large-scale accretion scenario for the formation of stellar clusters.

\section{Conclusions}
\label{sec_conclusions}

We analyzed sensitive VLA D-configuration (beam $\sim$3\farcs5) observations of the  $(1,1)$ and $(2,2)$ inversion transitions of ammonia, and optical \halpha{} and \sii{} emission in L1287. 
\begin{itemize}

\item 
The \sii{} emission, with a distribution similar to that of the \halpha{}, shows a strong peak toward the position of RNO1. There is also strong \sii{} emission coming from a compact ($\sim10''$, or $\sim$9\,000 au) region which encompasses the positions of the FU Ori binary RNO 1B/C, while the positions of IRAS/VLA 3, RNO 1F, RNO 1G/VLA 2, and VLA 4 fall in the periphery, where the emission is weaker. The \sii{} emission is probably tracing the shocks at the base of the jet driving the molecular outflow. 

\item The \nh{} emission reveals a main elongated structure, centered near the positions of the FU Ori binary RNO 1B/C and its associated still-forming cluster of YSOs. 
This ammonia structure extends over the observed field of view in the northwest-southeast direction, coinciding with the direction of the dust filament observed by \Herschel{} and perpendicularly to the bipolar molecular outflow in the region. 
A secondary ammonia core is observed $\sim40''$ ($\sim0.17$ pc) from the filaments, toward the position of RNO1, near the edge of our field of view.

\item The \nh{} emission is clumpy, with the channel map at $-18$ \kms{} revealing three ammonia clumps apparently associated with the YSOs RNO 1B, RNO 2C/VLA 1, and IRAS/VLA 3. 
These ammonia clumps coincide with the positions of three of the 1.3 mm continuum dust clumps identified by \citet{Jua19}, who infer a very high degree of fragmentation in this central region.

\item 
Three different velocity components were identified in the main ammonia structure. These components are referred to as the Guitar Core, the Blue Filament (including the RNO1 core), and the Red Filament,
in order of increasing velocity.

\item The Guitar Core is compact ($26''$, 0.12 pc in diameter) has a mass of 30 \Msol, and shows no signs of interaction with the embedded YSOs traced by the IR, mm, or cm continuum. 
It is probably seen in projection at the position of the YSOs.
This core shows a clear central-blue-spot infall hallmark. 
The central velocity of the \nh{} $(1,1)$ and $(2,2)$ lines is blueshifted $0.45$ \kms\ with respect to the velocity of the core ambient gas. 
This compact blueshifted spot at the center of the core is interpreted as infall onto a central protostar of 2.1 \Msol.

\item The Blue and Red Filaments are long ($\sim1100''$, 0.5 pc) and clearly trace the direction of the large-scale dust filament imaged by \Herschel{}. 
The molecular filaments traced by \nh{} have masses of 104 \Msol\ (Blue Filament) and 13 \Msol\ (Red Filament). 
The small-scale motions of the filament gas are nearly subsonic, except at the positions of the embedded YSOs.

\item The source VLA3 (IRAS 00338+6312), likely the most deeply embedded YSO and best candidate to drive the molecular outflow, appears directly associated with ammonia emission from the Red Filament, and with signs of dynamical perturbation, traced by an increased line width and local heating. 

\item The ammonia images, after the velocity component associated with the Guitar Core has been removed, reveal a cavity, with a diameter of $\sim12\,000$ au, towards the central region of the Blue/Red Filaments, with RNO 1C/VLA 1 located at its center, and several other sources located near its inner wall. 
The existence of such a cavity had already been proposed from near-IR and CS data.
The shape of the high-velocity CO outflow fits very well with the morphology of this cavity. 
The gas of the filaments shows a notable increase in velocity dispersion and rotational temperature at the interface between the outflow and the filaments, at the cavity walls near the positions of some of the associated YSOs. 
These results suggest that the cavity has been created by the interaction of the outflow with the dense gas of the filaments.

\item It has been possible to estimate the column densities of the filaments up to a distance of $\sim0.03$ pc from the filament axes. 
The column density across both filaments shows a marked decrease with distance to the filament axes. 
The mass per unit length of the filaments obtained from the integration of the observed column densities was $\sim60$ \Msolpc{}.
The fit with an isothermal cylindrical model gives a characteristic radius of 0.03 pc and a mass per unit length of $\sim150$ \Msolpc{}. 
The isothermal model predicts a velocity dispersion higher than that measured at small scales ($\sim0.03$ pc), consistent with the measured velocity gradient across the filament, and a the total mass per unit length of the filament of the order of twice the mass up to the characteristic radius, consistent with our results. 

\item The velocity gradient observed across the Blue Filament can be interpreted as infalling material onto the filament or, alternatively as rotation. 
The velocity gradients along the filaments are interpreted as inflow motions towards the location of the embedded YSOs. 

\item Additional ammonia observations with both higher angular and spectral resolution would be highly valuable to separate more clearly the three velocity components and better disentangle the kinematics and the role of the different embedded objects in the interaction between young stars, outflows, and dense gas in this region.

\end{itemize}

\begin{acknowledgements} 

We thank Eugenio Schisano for providing us with the \emph{Herschel} catalog of the region.
We thank the referee for his/her careful and helpful review of the paper.
This work has been partially supported by the 
State Agency for Research (AEI) of the Spanish MCIU through the 
AYA2017-84390-C2 grant (cofunded with FEDER funds), 
and through the
“Unit of Excellence Mar\'{\i}a de Maeztu 2020-2023” award to the Institute of Cosmos Sciences (CEX2019-000918-M)
and the
`Center of Excellence Severo Ochoa' award for the Instituto de Astrof\'{\i}sica de Andaluc\'{\i}a (SEV-2017-0709).
A. P. acknowledges financial support from CONACyT and UNAM-PAPIIT IN113119 grant, M\'exico.

\end{acknowledgements}

{}

\begin{appendix}

\section{\nh{} channel maps, spectra, and additional maps}
\label{achanmaps}

\begin{figure}[htb]
\resizebox{\hsize}{!}{\includegraphics{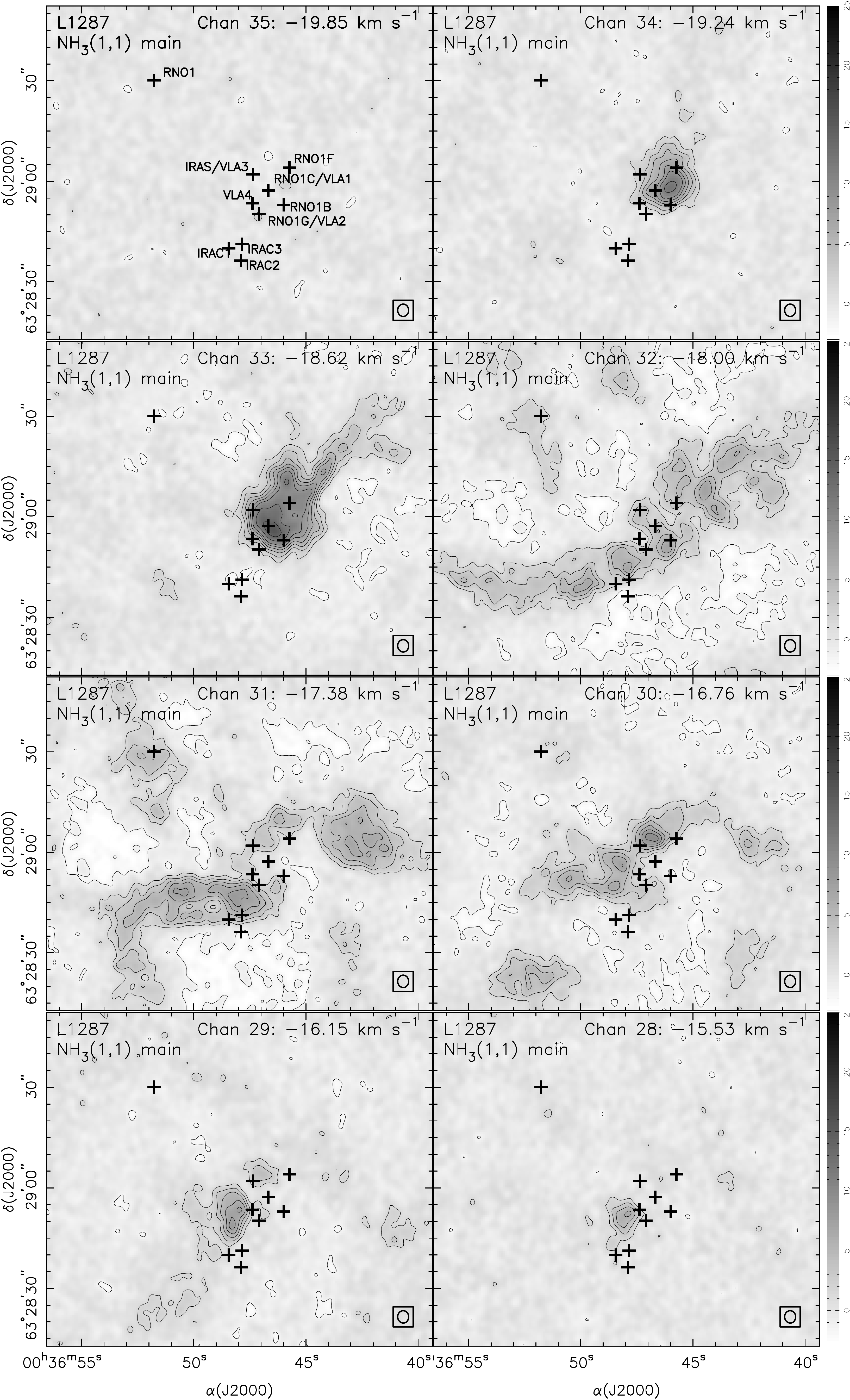}}
\caption{
Channel maps of the \nh{} $(1,1)$ main line intensity, not corrected for primary beam response. 
Contours are $-3$, 3, 6, 9, 12, 15, 18, 23 and 28 times 0.5 K, the rms of the maps.
The crosses correspond to YSOs identified in Fig.\ \ref{caha}
}
\label{fchanmap11}
\end{figure}

\begin{figure}[htb]
\resizebox{\hsize}{!}{\includegraphics{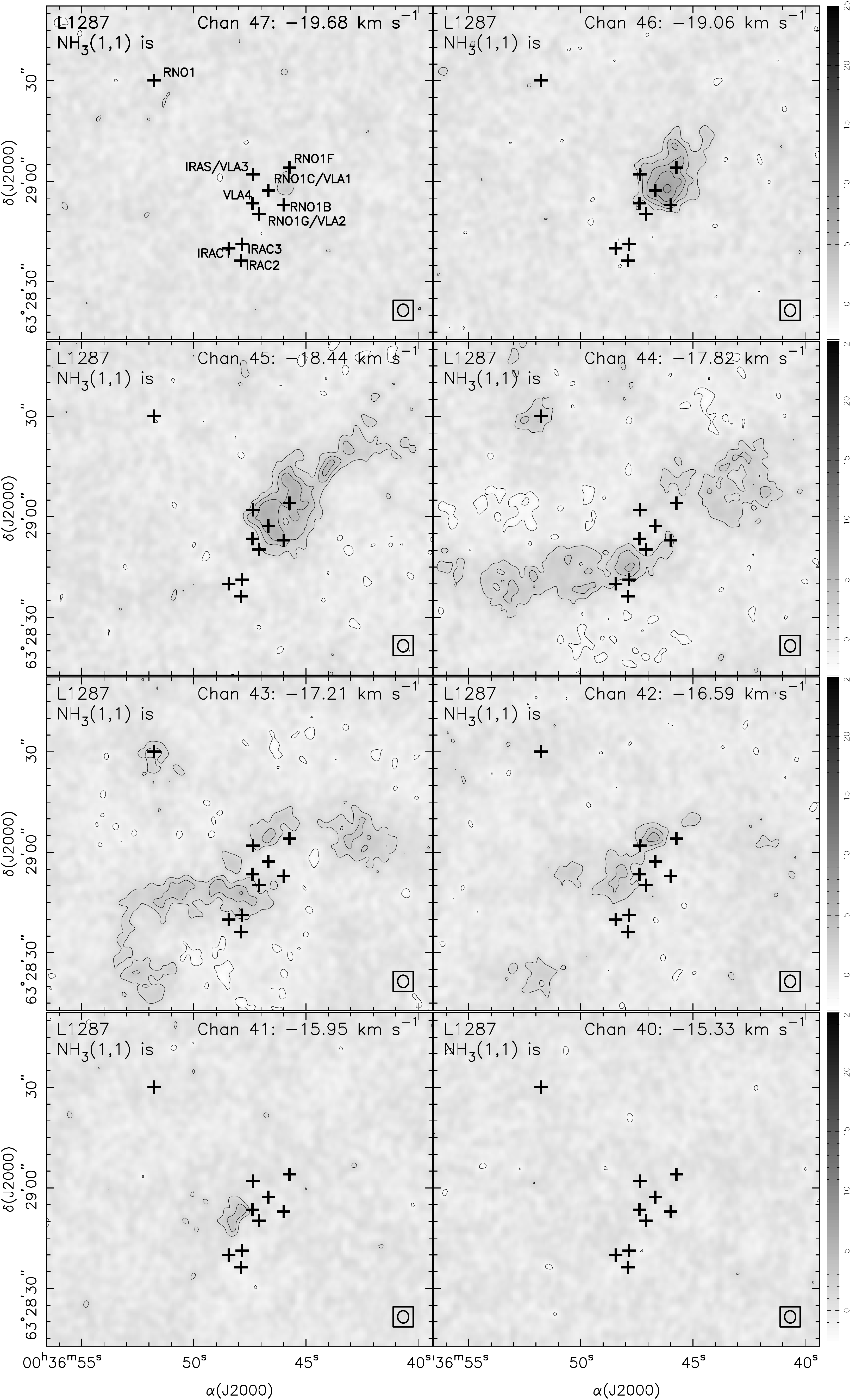}}
\caption{
Channel maps of the \nh{} $(1,1)$ inner-satellite line intensity, not corrected for primary beam response. 
Contours are $-3$, 3, 6, 9, 12, 15, 18, 23 and 28 times 0.5 K, the rms of the maps.
The crosses correspond to YSOs identified in Fig.\ \ref{caha}
}
\label{fchanmap11sat}
\end{figure}

\begin{figure}[htb]
\resizebox{\hsize}{!}{\includegraphics{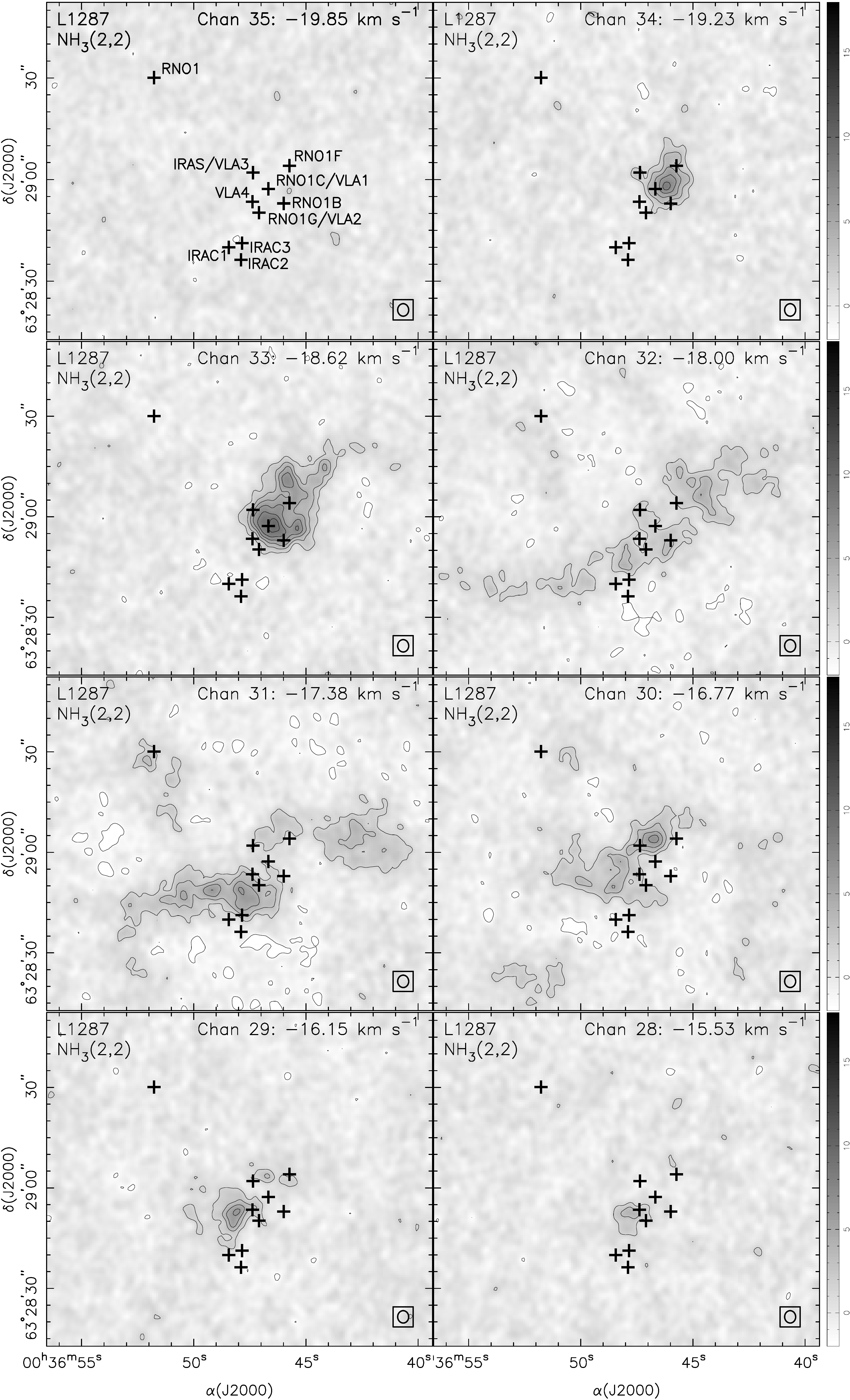}}
\caption{
Channel maps of the \nh{} $(2,2)$ main line intensity, not corrected for primary beam response.
Contours are $-3$, 3, 6, 9, 12, 15, and 18 times 0.5 K, the rms of the maps.
The crosses correspond to YSOs identified in Fig.\ \ref{caha}
}
\label{fchanmap22}
\end{figure}

\begin{figure}[htb]
\resizebox{\hsize}{!}{\includegraphics{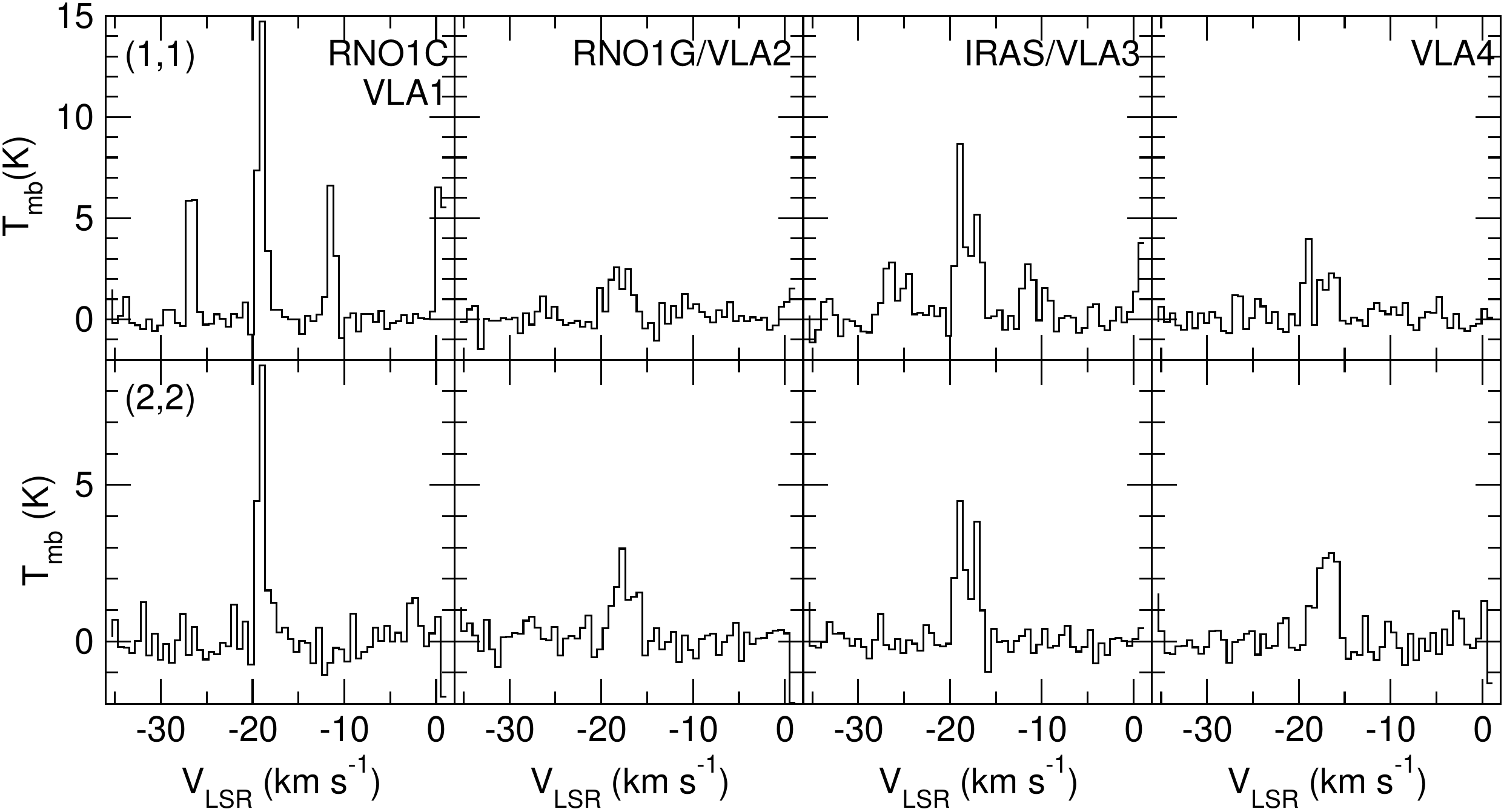}}
\resizebox{\hsize}{!}{\includegraphics{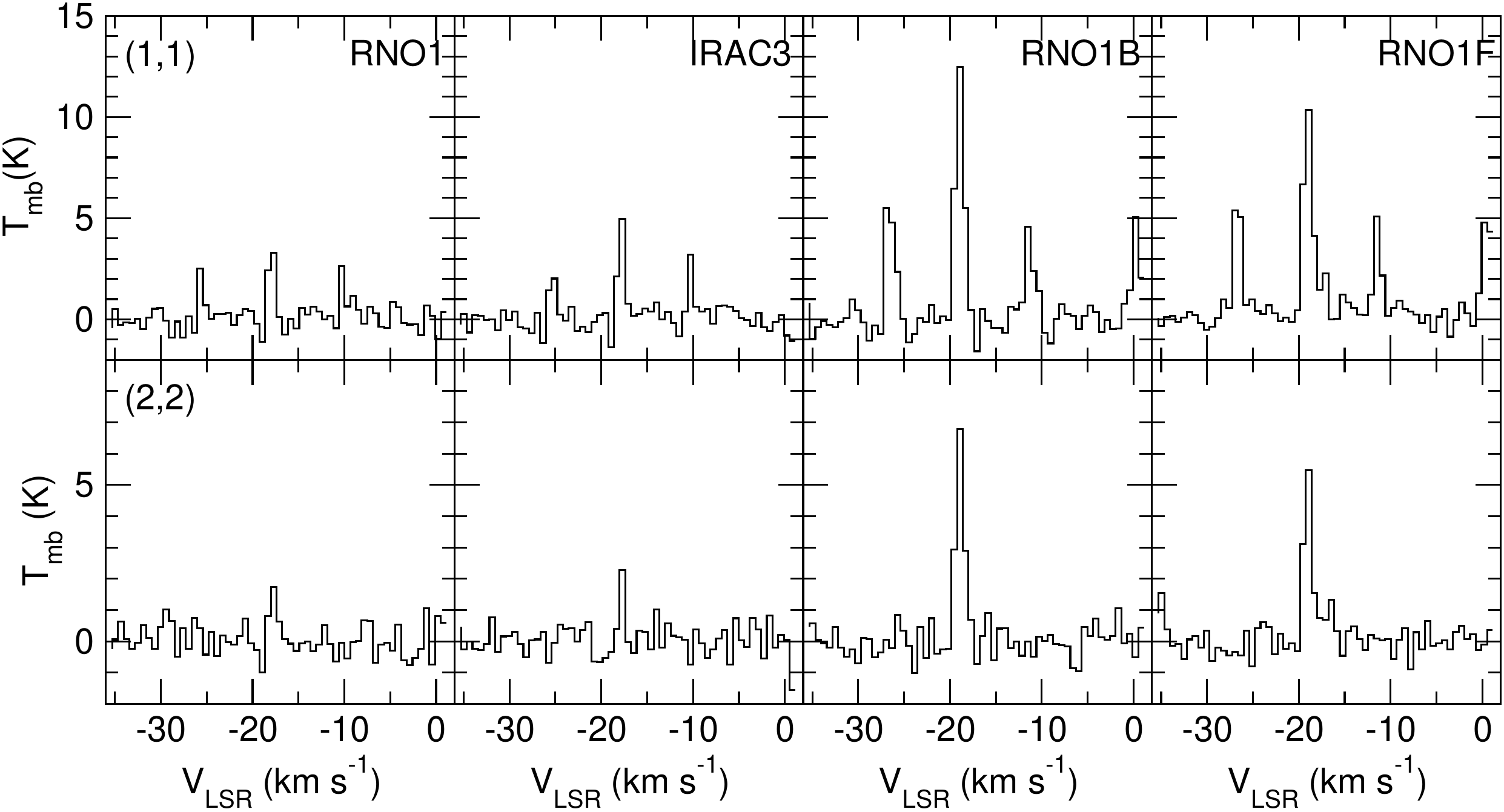}}
\caption{
\nh{} $(1,1)$ and $(2,2)$ spectra at the position of the embedded centimeter continuum and IR sources.
}
\label{fspectra_vla}
\label{fspectra_ir}
\end{figure}

\begin{figure}[htb]
\resizebox{\hsize}{!}{\includegraphics{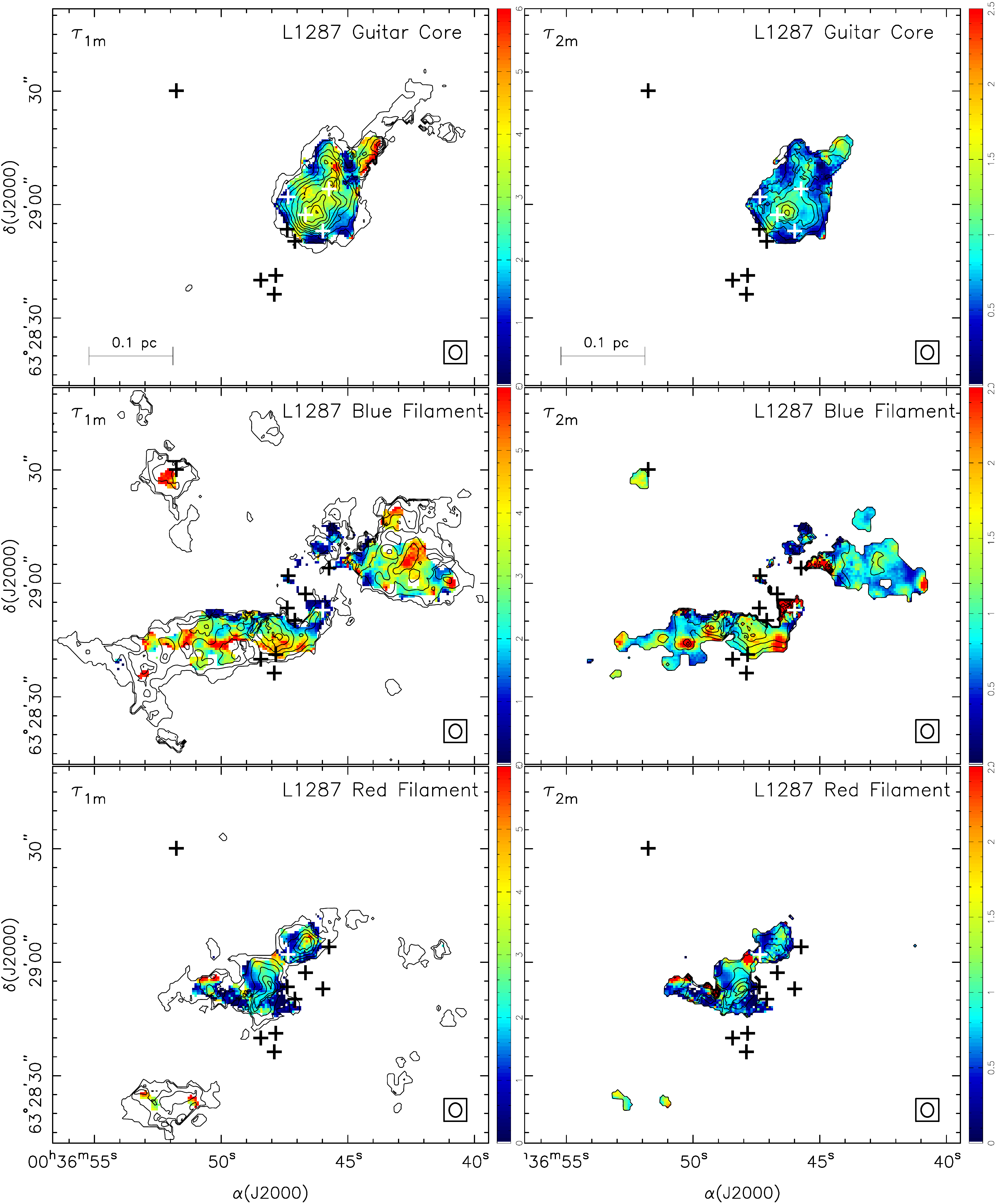}}
\caption{
L1287 \nh{} $(1,1)$ (left) and $(2,2)$ (right) main line optical depths, $\tau_{1m}$ and $\tau_{2m}$.
The color scale ranges from 0 to 6 for $\tau_{1m}$ and from 0 to 2.5 for $\tau_{2m}$. 
The contours are the $(1,1)$ (left) and $(2,2)$ (right) integrated line intensities.
The crosses correspond to the YSOs identified in Fig.\ \ref{caha}.
}
\label{ftau12}
\end{figure}

In this Appendix we show additional maps obtained from the analysis of the \nh{} $(1,1)$ and $(2,2)$ emission in L1287. 

The channel maps of the \nh{} $(1,1)$ and $(2,2)$ line intensity, not corrected for primary beam response, are shown for the 
\nh{} $(1,1)$ main line            (Fig.\ \ref{fchanmap11}),
\nh{} $(1,1)$ inner-satellite line (Fig.\ \ref{fchanmap11sat})
and
\nh{} $(2,2)$ main line            (Fig.\ \ref{fchanmap22}). 

The observed \nh{} $(1,1)$ and $(2,2)$ spectra at the position of the embedded centimeter continuum and IR sources are shown in
Fig.\ \ref{fspectra_vla}.

Finally, the \nh{} $(1,1)$ and $(2,2)$  main line optical depths, 
$\tau_{1m}$ and $\tau_{2m}$, 
are shown in Fig.\ \ref{ftau12}.

\section{Self-gravitating isothermal cylindrical filament model}
\label{amodel}

The case of a self-gravitating isothermal cylindrical filament of infinite
length was studied by \citet{Ost64} and used by \citet{Arz11, Arz13} in the analysis of filaments.
The cylindrical filament is characterized by its mass density as a function of the distance to the cylinder axis, $\rho(r)$, and by its temperature $T$, assumed to be uniform.
\citet{Ost64} derives an analytical solution for the equation of static equilibrium, with the boundary conditions imposed, i.e.\ that the density at the cylinder axis is finite,  $\rho(0)=\rho_0$, and derivable, $d\rho/dr(0)=0$. The solution can be written as
\begin{equation} 
\rho= \frac{\rho_0}{[1+(r/h)^2]^2}, 
\end{equation}
where $h$ is given by
\begin{equation} 
h^2= \frac{2 c_s^2}{\pi G \rho_0},
\end{equation}
and $c_s$ is the isothermal sound speed,
\begin{equation}
c_s= \left(\frac{kT}{\mu_\mathrm{gas} m_\mathrm{H}}\right)^{1/2}, 
\end{equation}
with 
$\mu_\mathrm{gas}$ being the mean molecular mass of the interstellar molecular gas, $\mu_\mathrm{gas}=2.3$ for a 10\% He abundance, and 
$m_\mathrm{H}$ the mass of the hydrogen atom.
This solution depends only on one free parameter, the density at the center of the cylinder, $\rho_0$. 
The parameter $h$ gives the characteristic radius of the
filament, for which the density drops to $\rho_0/4$.
The density given by this solution is flat at the center of the filament, and decreases sharply outwards as a power law with a power-law index $-4$,
\begin{equation}
\left\{\begin{array}{llll}
\rho(r) &\simeq& \rho_0         & (r\ll h), \\
\rho(h) &=     & \rho_0/4,                 \\
\rho(r) &\simeq& \rho_0 \,r^{-4}  & (r\gg h). 
\end{array}\right.
\end{equation}

The mass per unit length of the filament, up to a radius $r$, $\Ml(r)$, 
can be obtained by integration of the density, 
\begin{equation} 
\Ml(r) = 2\pi\int_0^r \rho(r)\,r\,dr=   \frac{\pi\rho_0 h^2}{1+(h/r)^2}=
\frac{2 c_s^2/G}{1+(h/r)^2}. 
\end{equation}
The total mass per unit length, $\Ml\equiv \Ml(r\to\infty)$, remains finite, 
\begin{equation} 
\Ml= \pi\rho_0 h^2= \frac{2 c_s^2}{G}, 
\end{equation}
and depends only on the temperature of the gas (through the value of $c_s$).
One half of the total mass is within a radius $r=h$, 
\begin{equation} 
\Ml(r=h)= \frac{1}{2}\Ml. 
\end{equation}

The relation between the mass density of the gas, $\rho$, and the molecular hydrogen number density, $n$, is given by
\begin{equation}
\rho= \mu_\mathrm{H_2} m_\mathrm{H} n,
\end{equation}
where $\mu_\mathrm{H_2}$ is the mean molecular gas per hydrogen molecule, 
$\mu_\mathrm{H_2}=2.8$ for a 10\% He abundance.

When the filament is observed perpendicularly, i.e.\ the line-of-sight is
perpendicular to the cylinder axis, the column density will depend only on the
projected radial distance from the filament axis, $p$, and will be
\begin{equation} 
N(p)= 2 \int_0^\infty n(r)\,dz, 
\end{equation}
where $z$ is the coordinate along the line of sight, and $p^2+z^2= r^2$.
By changing the integration variable from $z$ to $r$, with 
$dz= r\,dr/\sqrt{r^2-p^2}$, we obtain
\begin{equation} 
N(p)= 2 \int_p^\infty n(r)\frac{r\,dr}{\sqrt{r^2-p^2}}, 
\end{equation}
which is the standard form of the Abel Transform of $n(r)$ \citep{Bra00}.
The integral can be calculated analytically and the result is
\begin{equation}\label{eqnp}
N(p) = \frac{\pi}{2} n_0 h \left(1+\frac{p^2}{h^2}\right)^{-3/2}. 
\end{equation}
The column density of the isothermal cylindrical filament is flat at the center
of the filament, and decreases as a power law outwards with a power-law index
$-3$, 
\begin{equation}
\left\{\begin{array}{lll}
N(p) \simeq N_0 = \displaystyle\frac{\pi}{2} n_0 h    &\qquad& (p\ll h), \\ 
N(h) = 2^{-3/2} N_0 = 0.354 \, N_0\\
N(p) \simeq N_0 \displaystyle\left(\frac{p}{h}\right)^{-3} &\qquad& (p\gg h).
\end{array}\right.
\end{equation}

The model depends on two parameters that can be fitted to the observational data,
$N_0$, the column density at the filament axis, and
$h$, the characteristic radius of the filament. 
The expressions of the other physical parameters of the filament can be obtained from $N_0$ and $h$.
The expressions for the gas temperature $T$, the total mass per unit length $M_\mathrm{lin}$, and the number density at the filament axis, $n_0$, are
\begin{eqnarray}
T \equal 
\frac{2\mu_\mathrm{gas} \, \mu_\mathrm{H_2} \, m_\mathrm{H}^2}{k} N_0 \, h,\nonumber\\
M_\mathrm{lin} \equal 
(2\mu_\mathrm{H_2} \, m_\mathrm{H}) \, N_0 \, h,\\
n_0 \equal 
\frac{2}{\pi} N_0 \, h^{-1}.\nonumber
\end{eqnarray}
In practical units, the expressions become
\begin{eqnarray} 
\left[\frac{T}{\mathrm{K}}\right] \equal
2.69
\left[\frac{N_0}{10^{20}\ \mathrm{cm^{-2}}}\right]
\left[\frac{h}{\mathrm{pc}}\right],\nonumber\\
\left[\frac{M_\mathrm{lin}}{M_\odot\mathrm{~pc}^{-1}}\right] \equal 
4.49
\left[\frac{N_0}{10^{20}\ \mathrm{cm^{-2}}}\right] 
\left[\frac{h}{\mathrm{pc}}\right],\\
\left[\frac{n_0}{\mathrm{cm^{-3}}}\right] \equal
20.6
\left[\frac{N_0}{10^{20}\ \mathrm{cm^{-2}}}\right]
\left[\frac{h}{\mathrm{pc}}\right]^{-1}.\nonumber
\end{eqnarray}

\section{Free-fall of a cylindrical filament}
\label{atff}

The free-fall of a cylindrical filament of infinite length was studied, among others, by \citet{Pen69}, 
while the more general case of a finite length filament has also been object of several studies \citep[see for instance][and references therein]{Pon12}. 

In the following we will summarize the results of \citet{Pen69} and give the expressions used to analize the filaments in L1287.

For a cylindrical mass distribution, the gravitational field is radial (perpendicular to the cylinder axis). The gravitational acceleration is
\begin{equation}
\label{eqgcyl}
g(r)= 2G \frac{M_\mathrm{lin}(r)}{r},
\end{equation}
where $M_\mathrm{lin}(r)$ is the mass per unit length of the cylinder, inside a radius $r$. 
The mass for radii greater than the distance $r$ to the axis does not contribute to the gravitational force. 
The former expression can be written as a differential equation in the radial
velocity $v$,
\begin{equation}
    v\,dv= -2GM_\mathrm{lin}(a)\frac{dr}{r},
\end{equation}
where $M_\mathrm{lin}(a)$ is the mass per length unit at the initial radius of the free
fall, $a$, a value of the mass that is constant during the free fall. The integration of the
differential equation from the initial radius, $a$, with zero initial velocity, gives
\begin{equation}\label{eqc3}
    v^2= 4GM_\mathrm{lin}(a)\ln\frac{a}{r}
\end{equation}
It is useful to define the characteristic velocity $v_c$ ($\sqrt{2}$ times the Keplerian orbital velocity),
\begin{equation}\label{eqc4}
    v_c= \sqrt{4GM_\mathrm{lin}(a)},
\end{equation}
so that
\begin{equation}\label{eqc5}
    v^2= {v_c}^2 \ln\frac{a}{r}.
\end{equation}
This is a differential equation in the radius $r(t)$,
\begin{equation}\label{eqc6}
    \frac{dr}{dt}= -v_c\left(\ln\frac{a}{r}\right)^{1/2},
\end{equation}
whose integration, with the initial condition $r(0)=a$, can be given as
\begin{equation}\label{eqc7}
    v_c\,t = -\int_{a}^r
    \frac{dx}{\left(\ln\displaystyle\frac{a}{x}\right)^{1/2}}.
\end{equation}
By using the adimensional parameter $\theta$ \citep{Pen69}, 
\begin{equation}\label{eqc8}
  \theta= \left(\ln\displaystyle\frac{a}{r}\right)^{1/2},
\end{equation}
the integral of Eq.\ \ref{eqc7} results in
\begin{equation}\label{eqc9}
    t = \sqrt{\pi} \frac{a}{v_c} \, \mathrm{erf}\,{\theta},
\end{equation}
where erf is the error function.

The free-fall time, $t_\mathrm{ff}$, is obtained for $r\to0$, implying $\theta\to\infty$ (Eq.\ \ref{eqc8}), so that  erf$\,\theta\to1$, and Eq.\ \ref{eqc9} gives
\begin{equation}\label{eqc10}
    t_\mathrm{ff} = \sqrt{\pi} \frac{a}{v_c} = \left(\frac{\pi a^2}{4GM_\mathrm{lin}(a)}\right)^{1/2}.
\end{equation}

In summary, the equations of the collapse can be written as a function of the parameter $\theta$, which varies from 0 (beginning of the collapse, $r=a$), to $\infty$ (end of the collapse, $r\to0$),
\begin{equation}\label{eqc11}
    \left\{
    \begin{array}{l}
    t = t_\mathrm{ff} \, \mathrm{erf}\,{\theta}, \\
    r = a \exp(-\theta^2),\\
    v = -v_c \, \theta,
    \end{array}
    \right.
\end{equation}
where 
$a$ is the initial radius, 
$v_c$ is given by Eq.\ \ref{eqc4}, and 
$t_\mathrm{ff}$ by Eq.\ \ref{eqc10}.

In practical units, the characteristic velocity $v_c$ (Eq.{} \ref{eqc4}) is given by
\begin{equation}\label{eqc12}
  \left[\frac{v_c}{\mathrm{km~s^{-1}}}\right] =
  0.131
  \left[\frac{M_\mathrm{lin}(a)}{M_\odot\mathrm{~pc}^{-1}}\right]^{1/2},
\end{equation}
and the free-fall time from an initial radius $a$ (Eq.{} \ref{eqc10}) is
\begin{equation}\label{eqc13}
  \left[\frac{t_\mathrm{ff}}{\mathrm{Myr}}\right] =
  13.2
  \left[\frac{a}{\mathrm{pc}}\right]
  \left[\frac{M_\mathrm{lin}(a)}{M_\odot\mathrm{~pc}^{-1}}\right]^{-1/2}
\end{equation}

Let us now assume that we observe a free-fall velocity $v$ at a distance $r$ from the filament axis, and we want to know from which radius, $a$, the free-fall initiated, and the free-fall time. In addition, let us assume that the radius of observation $r$ is $r\gg h$ (that is, the infall velocity is observed far enough from the filament axis), so that $M_\mathrm{lin}(a)$ is a good approximation to the total mass of the filament $M_\mathrm{lin}$. By using Eq.{} \ref{eqc11}, the initial radius of the infall is
\begin{equation}\label{eqc14}
    a = r \exp\left(\frac{v^2}{{v_c}^2}\right),
\end{equation}
which allows us to check whether $a\gg h$, as assumed. 
The total free-fall time from the initial radius $a$ is
\begin{equation}\label{eqc15}
  t_\mathrm{ff}= \sqrt{\pi} \frac{r}{v_c} \exp\left(\frac{v^2}{{v_c}^2}\right),
\end{equation}
In practical units, the initial radius $a$ (Eqs.{} \ref{eqc12}, \ref{eqc4}) is given by
\begin{equation}\label{eqc16}
 \left[\frac{a}{\mathrm{pc}}\right] =
  \left[\frac{r}{\mathrm{pc}}\right]
  \exp\left(
  58.1
  \left[\frac{v}{\mathrm{km~s^{-1}}}\right]^2
  \left[\frac{M_\mathrm{lin}}{M_\odot\mathrm{~pc}^{-1}}\right]^{-1}
  \right), 
\end{equation}
and the total free-fall time $t_\mathrm{ff}$ (Eqs.{} \ref{eqc13}, \ref{eqc4}) is
\begin{eqnarray}\label{eqc17}
  \left[\frac{t_\mathrm{ff}}{\mathrm{Myr}}\right] &\!\!=\!\!&
  13.2
  \left[\frac{r}{\mathrm{pc}}\right]
  \left[\frac{M_\mathrm{lin}}{M_\odot\mathrm{~pc}^{-1}}\right]^{-1/2} \times \nonumber\\
  &&
  \exp\left(
  58.1
  \left[\frac{v}{\mathrm{km~s^{-1}}}\right]^2
  \left[\frac{M_\mathrm{lin}}{M_\odot\mathrm{~pc}^{-1}}\right]^{-1}
  \right).
\end{eqnarray}

\end{appendix}

\end{document}